\documentclass[aps,amsmath,amssymb,reprint,floatfix,longbibliography,prapplied]{revtex4-1}
\usepackage{graphicx}
\usepackage{dcolumn}
\usepackage{bm}
\usepackage[pdfencoding=auto, psdextra]{hyperref}
\usepackage{comment}
\usepackage[version=4]{mhchem}
\usepackage{color}
\usepackage{listings}

\usepackage{siunitx}
\DeclareSIPostPower\tothefourth{4}
\usepackage[english]{babel}
\usepackage{esdiff} 
\usepackage{mathtools}
\DeclarePairedDelimiter\abs{\lvert}{\rvert} 
\addto\captionsenglish{}
\addto\captionsenglish{}

\begin{document}
\title{Current detection using a Josephson parametric upconverter}
\author{Felix E. Schmidt}
\author{Daniel Bothner}
\author{Ines C. Rodrigues}
\author{Mario F. Gely}
\author{Mark D. Jenkins}
\author{Gary A. Steele}\email[]{g.a.steele@tudelft.nl}
\affiliation{Kavli Institute of NanoScience, Delft University of Technology, Lorentzweg 1, 2628 CJ, Delft, The Netherlands.}


\begin{abstract}

We present the design, measurement and analysis of a current sensor based on a process of Josephson parametric upconversion in a superconducting microwave cavity. 
Terminating a coplanar waveguide with a nanobridge constriction Josephson junction, we observe modulation sidebands from the cavity that enable highly sensitive, frequency-multiplexed output of small currents for applications such as transition-edge sensor array readout. 
We derive an analytical model to reproduce the measurements over a wide range of bias currents, detunings and input powers.
Tuning the frequency of the cavity by more than \SI{100}{\mega\hertz} with DC current, our device achieves a minimum current sensitivity of \SI{8.9}{\pico\ampere\per\sqrt{\hertz}}.
Extrapolating the results of our analytical model, we predict an improved device based on our platform, capable of achieving sensitivities down to \SI{50}{\femto\ampere\per\sqrt{\hertz}}, or even lower if one could take advantage of parametric amplification in the Josephson cavity.
Taking advantage of the Josephson architecture, our approach can provide higher sensitivity than kinetic inductance designs, and potentially enables detection of currents ultimately limited by quantum noise.
\end{abstract}

\maketitle

\section{Introduction}

Ultra-low noise radiation detection has applications in astronomy, particle physics, and quantum information processing.
In particular, transition edge sensors (TES) allow for broadband radiation detection with exceptionally low noise equivalent power~\cite{goldieUltralownoiseMoCuTransition2011} and photon number resolution~\cite{cabreraDetectionSingleInfrared1998,millerDemonstrationLownoiseNearinfrared2003}.
To read out the small changes in current of TES in response to radiation absorption, highly sensitive current amplifiers such as superconducting quantum interference devices (SQUIDs) can be used with sensitivities as low as \SI{4}{\femto\ampere\per\sqrt{\hertz}}~\cite{gayUltralowNoiseCurrent2000}.
However, with the increasing number of TES to be read out simultaneously in multipixel detectors, SQUID amplifiers significantly increase system cost and complexity, especially when employing frequency-domain multiplexing to reduce the number of necessary amplifiers~\cite{hendersonReadoutTwokilopixelTransitionedge2016}.

An example of recently developed current detectors as a replacement of SQUIDs are kinetic inductance parametric upconverters (KPUPs), also referred to as microwave kinetic inductance nanowire galvanometers, which rely on the changing kinetic inductance $L_k$ of a narrow superconducting wire embedded in a microwave circuit in response to a DC bias current, with state of the art devices reaching current sensitivities $\mathcal{S}_I$ between \SIrange{5}{10}{\pico\ampere\per\sqrt{\hertz}}~\cite{kherKineticInductanceParametric2016,doernerCompactMicrowaveKinetic2018,kuzminTerahertzTransitionEdgeSensor2018}.
One could potentially achieve a higher response from such a cavity detector by replacing the nanowire kinetic inductance element with a Josephson junction (JJ), enabling detection of currents using a Jospheson parametric upconverter (JPUP). 
This would also enable the incorporation of processes such as Josephson parametric amplification, which allows signals to be amplfied with quantum limited noise~\cite{stehlikFastChargeSensing2015}, directly in the readout cavity.

Typically, the integration of JJs in superconducting microwave circuits is technologically more demanding due to the additionally needed fabrication steps to avoid aging effects and low coherence at microwave frequencies~\cite{pavolotskyAgingAnnealinginducedVariations2011,gotetiReliabilityStudiesNb2019,gunnarssonDielectricLossesMultilayer2013,yanaiObservationEnhancedCoherence2019}.
The intrinsically large Kerr-nonlinearity of JJs~\cite{wallraffStrongCouplingSingle2004} can additionally place an upper limit on the device power allowed for circuit operation, which calls for either large critical current JJs with additional fabrication challenges~\cite{lecocqJunctionFabricationShadow2011}, or appropriate circuit design for sufficiently diluting the nonlinearity to provide stable device operation.

Here, we provide experimental realisation of a JPUP based on a hybrid combination of a direct current (DC) accessible microwave cavity in coplanar waveguide (CPW) geometry~\cite{bosmanBroadbandArchitectureGalvanically2015a,schmidtBallisticGrapheneSuperconducting2018}.
The design uses a constriction JJ fabricated in the same step and layer as the microwave cavity which simplifies the fabrication procedure and allows for high cavity drive powers~\cite{vijayOptimizingAnharmonicityNanoscale2009a,kennedyTunableNbSuperconducting2019a,rodriguesCouplingMicrowavePhotons2019a,bothnerPhotonPressureStrongCouplingTwo2019}.
We show device operation by converting \si{\kilo\hertz} current signals to the \si{\giga\hertz} range, and reproduce the data with an analytical model for a wide range of bias currents, drive detunings and drive powers.
Our device achieves performance comparable to KPUP technology, with the potential to provide enhanced current sensitivity with a more optimized design. Ultimately, by using Josephson parametric amplification in the same cavity as used for sensing, the JPUP could sense low frequency currents with a sensitivity limited by quantum noise.


\begin{figure*}
	\centering
	\includegraphics[]{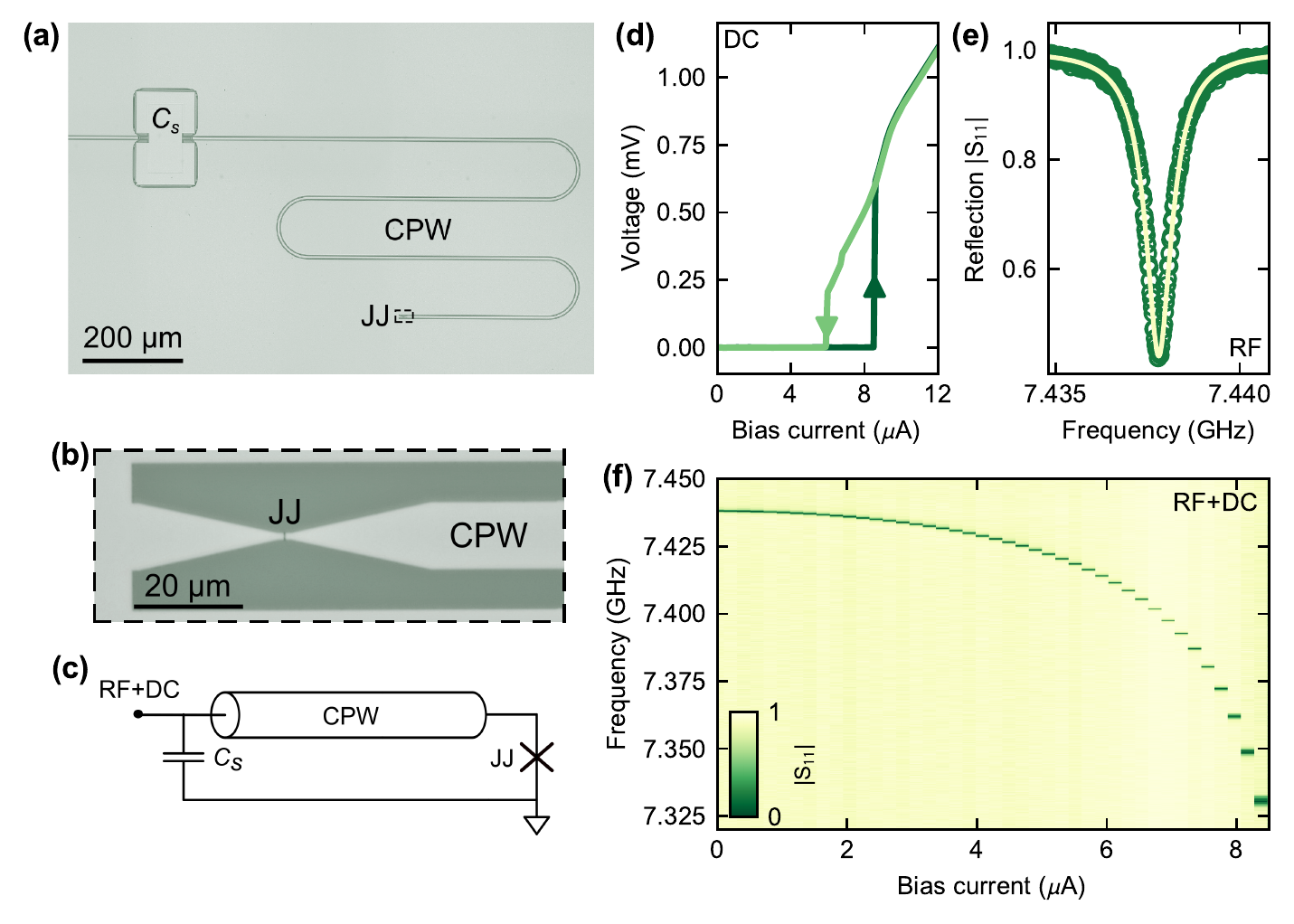}
	\caption{
		\textbf{A coplanar microwave Josephson circuit with direct current bias.}
		(a) Optical image of the measured device.
		It consists of a coplanar waveguide transmission line shunted to ground via a parallel plate capacitor $C_s$ on the input, and the Josephson junction shorting the CPW center conductor to ground on the far end.
		(b) Optical close-up of the area around the JJ.
		(c) Schematic circuit layout.
		(d) Current-voltage characteristics of the JJ, measured by sweeping the bias current up and down (sweep direction indicated by arrows).
		(e) Normalized and background-corrected reflection $\abs{S_{11}}$ of the device with zero bias current applied, cf. Sec.~\ref{sec:general-S11} of the Supplemental Material~\cite{SeeSupplementalMaterial}.
		Circles: data, line: fit.
		(f) Reflection coefficient $\abs{S_{11}}$ as a function of bias current.
		Since the Josephson inductance increases with bias current, the resonance frequency of the circuit shifts towards lower values.
		\label{fig:figure1}
	}
\end{figure*}

\section{The DC bias microwave circuit}

The device consists of a galvanically accessible microwave cavity, formed by a CPW that is shunted by an input capacitor $C_s$ and shorted to ground at its far end by a JJ, as depicted in Figs.~\ref{fig:figure1}(a-c).
The JJ is formed by a narrow constriction in the superconducting base-layer, which allows us to fabricate it in the same step as the microwave circuit.
For details on the fabrication procedure, see Sec.~\ref{sec:fabrication} of the Supplemental Material~\cite{SeeSupplementalMaterial}.
Due to the shunt capacitor allowing low-frequency signals to pass through, but acting as a semi-transparent mirror for microwave frequencies, our circuit allows for simultaneous measurements in the DC and RF regimes.

In the DC regime, the CPW center conductor acts as a long lead to the JJ, which we use to perform a current-voltage measurement to characterize the JJ.
Upon applying an increasing DC bias current, the JJ switches from the superconducting to the voltage state and back again at switching and retrapping currents $I_s\approx\SI{8.5}{\micro\ampere}$ and $I_r\approx\SI{6.1}{\micro\ampere}$, as shown in Fig.~\ref{fig:figure1}(d).
The observed hysteresis is most likely a combination of the capacitances of the CPW and shunt capacitor, and local heating in the junction area, cf. Refs.~\cite{tinkhamIntroductionSuperconductivity1996,skocpolSelfHeatingHotspots1974,hazraHysteresisSuperconductingShort2010,kumarReversibilitySuperconductingNb2015} and Sec.~\ref{sec:hysteresis} in the Supplemental Material~\cite{SeeSupplementalMaterial}.

In the RF regime, the JJ acts as a nonlinear inductor, with its inductance $L_J$ depending on the amount of bias current $I_b$ flowing through it, according to
\begin{align}
	L_J(I_b)= \frac{\Phi_0}{2\pi \sqrt{I_c^2-I_b^2}}\ ,
	\label{eq:Lj-of-I}
\end{align}
with $I_c$ the critical current and $\Phi_0$ the magnetic flux quantum.
For zero bias current, both the impedance of shunt capacitor and of the JJ are small compared to the characteristic impedance of the CPW, i.e. $\omega L_J,1/\omega C_s \ll Z_0$.
The CPW can thus host a fundamental half-wavelength ($\lambda/2$) mode with current antinodes at both ends.
When recording the reflected signal of the device using single-tone RF spectroscopy, the reflection signal shows a dip in the spectrum as seen in Fig.~\ref{fig:figure1}(e).
We fit the data using the reflection coefficient of our circuit,
\begin{align}
	S_{11} = \frac{\kappa_\text{e}-\kappa_\text{i}-2i\Delta}{\kappa_\text{e}+\kappa_\text{i}+2i\Delta}\ ,
	\label{eq:general:S11}
\end{align}
with $\Delta=\omega-\omega_0$ the detuning between a drive at $\omega$ and the resonance frequency $\omega_0$ and the external and internal loss rates $\kappa_\text{e}$ and $\kappa_\text{i}$, respectively.
At zero bias current, we find a resonance frequency of $\omega_0=2\pi\times\SI{7.438}{\giga\hertz}$, and linewidths of $\kappa_\text{e}=2\pi\times\SI{624}{\kilo\hertz}$ and $\kappa_\text{i}=2\pi\times\SI{261}{\kilo\hertz}$.

As we DC-bias the circuit, $L_J$ increases, effectively shifting the voltage antinode closer to the JJ.
This results in a continuously decreasing resonance frequency, tuning over approximately \SI{108}{\mega\hertz}, cf. Fig.~\ref{fig:figure1}(f).
We can approximate the bias current dependence of the cavity resonance frequency with a model describing a $\lambda/2$ CPW resonator terminated by a JJ via
\begin{align}
	\omega_0(I_b)=\frac{\omega_{\lambda/2}}{1 + L_J(I_b,I_c)/L_r}
	\label{eq:f0vsI}
\end{align}
with $\omega_{\lambda/2}$ the resonance frequency of the CPW directly shorted to ground and $L_r$ the total bare resonator inductance  (see Sec.~\ref{sec:resfit} of the Supplemental Material~\cite{SeeSupplementalMaterial} and Ref.~\cite{pogorzalekHystereticFluxResponse2017}).
We use this model to fit the measured resonance frequencies in Fig.~\ref{fig:figure2}(a), from which we extract $\omega_{\lambda/2}=2\pi\times\SI{7.515}{\giga\hertz}$, $L_r=\SI{3.458}{\nano\henry}$ and $I_c=\SI{9.176}{\micro\ampere}$.
The resonator inductance agrees with the value expected from our circuit design.
The critical current as inferred from the microwave measurement is approximately \SI{8}{\percent} larger than the DC switching current.
We suspect that current noise in the DC line leads to premature switching of the JJ in the IV measurements, resulting in $I_s<I_c$, as discussed in Sec.~\ref{sec:lossrates} of the Supplemental Material~\cite{SeeSupplementalMaterial} and Ref.~\cite{kautzNoiseaffectedIVCurves1990b}.
On the other hand, the RF measurement is sensitive to the Josephson inductance, from which we can infer the critical current in a less perturbative way.
We note that current-biasing a superconducting wire will also change its kinetic inductance $L_k$~\cite{annunziataTunableSuperconductingNanoinductors2010,vissersFrequencytunableSuperconductingResonators2015}.
However, while our device does possess a noticeable kinetic inductance fraction~\cite{gaoExperimentalStudyKinetic2006}, the changes in $L_k$ within the range of applied bias currents are negligible compared to $L_J$ and we thus attribute the resonance frequency shift completely to the latter, cf. Sec.~\ref{sec:Lk} of the Supplemental Material~\cite{SeeSupplementalMaterial}.


\section{Current detection by frequency up-conversion}
\begin{figure}
	\centering
	\includegraphics[]{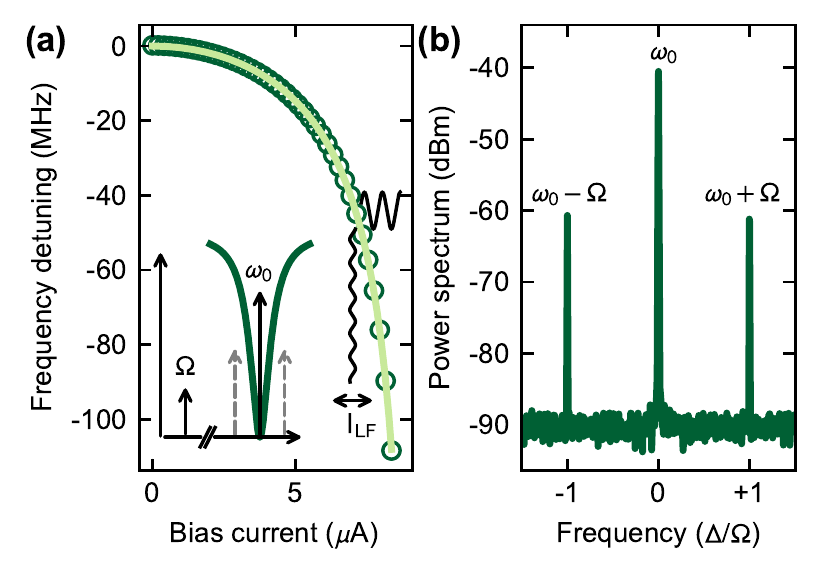}
	\caption{
		\textbf{Current detection by frequency up-conversion.}
		(a) Cavity resonance frequency for increasing DC bias current, showing a total frequency shift of $\SI{108}{\mega\hertz}$.
		Circles: measured data, line: fit to resonance frequency using Eq.\eqref{eq:f0vsI}.
		Inset: sketched measurement scheme in the frequency domain.
		By driving the cavity on resonance $\omega = \omega_0$ and simultaneously modulating with a low frequency current $\delta I = I_\text{LF}\cos\Omega t$, the cavity generates sidebands to the drive tone at $\omega_0 \pm \Omega$ (dashed grey arrows).
		(b) The power spectrum of the reflected field at $I_b=\SI{2.5}{\micro\ampere}$, containing the input pump signal at $\omega_0$ and the first order sidebands due to mixing at $\omega_0 \pm \Omega$.
		The noise floor sets a lower limit on the smallest detectable sideband amplitude.
		The sideband amplitude allows us to directly calibrate the noise floor and thus the sensitivity from the signal-to-noise ratio, here $\text{SNR}\approx\SI{30}{\decibel}$.
		\label{fig:figure2}
	}
\end{figure}

Figure~\ref{fig:figure2}(a) illustrates the principle of current detection using the DC biased Josephson cavity.
To detect small modulation currents, we drive the cavity on resonance $\omega_0(I_b)$ and simultaneously modulate the bias point $I_b$ with a low-frequency signal $\delta I=I_\text{LF}\cos{\Omega t}$, so that the total current is given by $I=I_b+I_\text{LF}\cos\Omega t$.
The responsivity of the resonance frequency to bias current,
\begin{align}
	G_1 = \frac{\partial\omega_0}{\partial I_b} \ ,
\end{align}
exceeds $2\pi\times\SI{100}{\mega\hertz\per\micro\ampere}$ for $I_b\gtrsim\SI{8}{\micro\ampere}$.
As a consequence, once the resonance frequency is modulated by $I_\text{LF}$, phase modulation leads to the generation of sidebands in the microwave drive tone reflection with $\omega = \omega_d \pm n \Omega$, where $n \in \mathbb{Z}$.
The reflected cavity field thus exhibits the drive tone together with the sidebands, as depicted in Fig.~\ref{fig:figure2}(b).

The general equation of motion for the amplitude field $\alpha$ of a harmonic high-$Q$ oscillator with small nonlinearity $\beta$, written in the frame rotating with the drive, is given by
\begin{align}
    \dot{\alpha} = \left[ -i \left( \Delta+\beta\abs{\alpha}^2 \right)-\frac{\kappa}{2} \right]\alpha + \sqrt{\kappa_\text{e}} S_\text{in}\ ,
    \label{eq:Duffing-EOM}
\end{align}
with $S_\text{in}$ the amplitude of the drive field in units of $\sqrt{\si{Photons\per\hertz}}$ at $\omega$, and $\beta$ a small nonlinearity~\cite{castellanos-beltranDevelopmentJosephsonParametric2010}.
We consider the case in which the cavity resonance frequency is a function of an additional current given by $I = I_b + \delta I=I_b+I_\text{LF}\cos\Omega t$, such that
\begin{align}
	\begin{split}
		\omega_0 &= \omega_0(I_b) + \sum_{m=1}^n \frac{\partial^m \omega_0}{\partial I^m}\delta I^m = \omega_I + \sum_{m=1}^n G_m \delta I^m \ .
		\label{eq:omega_Taylor}
	\end{split}
\end{align}
The resulting field amplitude of the first order sidebands appearing at $\omega_0\pm1\Omega$ is
\begin{align}
	\abs{S_{\pm 1}}^2 = \frac{\kappa_\text{e}\alpha_0^2 G_1^2 I_\text{LF}^2}{\kappa^2+4(\Delta\pm\Omega)^2}\ .
	\label{eq:first_amplitude}
\end{align}
In our experiment, we chose $\Omega=2\pi\times\SI{1}{\kilo\hertz}$ and $I_\text{LF}=\SI{10}{\nano\ampere}$.
In this case, $\Omega\ll\kappa$ and red ($S_{-}$) and blue ($S_{+}$) sidebands have approximately equal amplitudes, see Sec.~\ref{sec:bluered} of the Supplemental Material~\cite{SeeSupplementalMaterial}.
Note that even higher order contributions from the current still contribute to the $\pm1\Omega$ sideband, but those contributions can be neglected for relatively weak modulation.

\begin{figure*}[]
	\centering
	\includegraphics[]{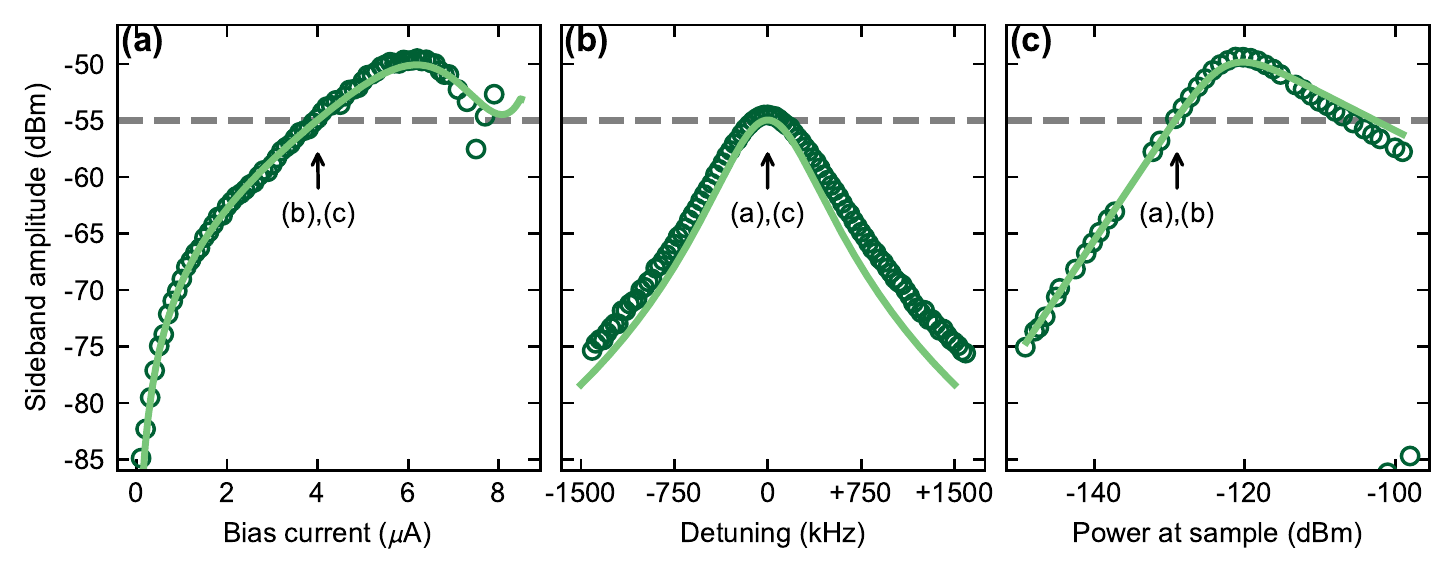}
	\caption{
		\textbf{Exploring the parameter space for the first order sideband amplitude.}
		(a) Sideband height for changing bias current setpoint at fixed input power and zero detuning. 
		(b) Sideband height for changing drive detuning at fixed input power and bias current.
		(c) Sideband height for changing input power at fixed bias current and detuning. 
		Circles: measured data, solid lines: calculated amplitude via input-output theory, dashed grey line: calculated sideband amplitude at $I_b=\SI{4}{\micro\ampere}$, $\Delta=0$ and $P_\text{in}=\SI{-129}{dBm}$.
		Arrows indicate the setpoints for the other respective panels.
		\label{fig:figure3}
	}
\end{figure*}

%
To explore the parameter space of our device, we performed a series of current-mixing measurements for different values of bias current $I_b$, drive detuning $\Delta$ and drive amplitude $S_\text{in}$, for all of which we observe excellent agreement between experiment and theory:
As can be seen in Fig.~\ref{fig:figure3}(a), for the case of varying bias current and as expected from Eqs.~\eqref{eq:f0vsI},\eqref{eq:first_amplitude}, the first order sideband vanishes for zero bias current.
As we increase the DC bias current, the increasing Josephson inductance leads to an increased responsivity $\partial\omega_0/\partial I_b$, which in turn results in a growing sideband amplitude.
Assuming all other parameters remain constant, the sideband amplitude should keep growing until the bias current reaches the critical current of the JJ, at which point the junction switches to the normal state, effectively destroying the device response.
However, already at $I_b \approx 0.75 I_c$ the sideband amplitude exhibits a maximum value and begins to decrease subsequently.
The origin for this phenomenon lies in the growth of $\kappa_\text{i}$ for increasing $I_b$, which limits the maximum achievable sideband amplitude, cf. Sec.\ref{subsec:limitations} and Sec.~\ref{sec:lossrates} of the Supplemental Material~\cite{SeeSupplementalMaterial}.

Operating the device at constant bias current and drive power $P_\text{in}$ but sweeping the drive tone with respect to the cavity resonance similarly reduces the sideband amplitude, which is reflected in both the theoretical model and our measurements, cf. Fig.~\ref{fig:figure3}(b).
We attribute deviations of the model from the data to an effectively increased cavity linewidth resulting from a noise-induced fluctuating cavity frequency.

Finally, when setting the detuning back to zero and sweeping the drive power, we initially observe a linear increase of the sideband amplitude, cf. Fig.~\ref{fig:figure3}(c).
This is in good agreement with the intracavity field dependence with pump power of a linear cavity.
However, due to the nonlinearity of the JJ and the resulting Kerr anharmonicity of the circuit, our device enters the Duffing regime for large input powers, resulting in the observable reduction of the sideband amplitude:
The anharmonicity results in a down shifted resonance frequency given by $\omega_0^\prime = \omega_0-\abs{\alpha_0}^2\beta$.
In the measurement depicted in Fig.~\ref{fig:figure3}(c), the only varying parameter is the pump power, which means that in the Duffing regime the drive acquires an increase in detuning for increased power, resulting in a decreased sideband amplitude, as we saw earlier.

\section{Current sensitivity}

Having established the validity of our theoretical framework, we calculate the current sensitivity $\mathcal{S}_I$ of our device.
This quantity captures the minimum current that the device is able to discriminate from the noise floor.
We obtain this quantity by extracting the signal-to-noise ratio (SNR) of the first sideband amplitude:
Since we know the amplitude of our ingoing LF current signal, we can convert the sideband amplitude and noise floor to currents as described in Sec.~\ref{sec:analysis} of the Supplemental Material~\cite{SeeSupplementalMaterial}.
We obtain
\begin{align}
	\mathcal{S}_I = \frac{I_\text{LF}}{\sqrt{\text{ENBW}\times10^{(S-N)/10}}}\ ,
	\label{eq:sensitivity}
\end{align}
with $\text{ENBW}$ the equivalent noise bandwidth of the spectrum analyzer~\cite{rauscherFundamentalsSpectrumAnalysis2016a}, and $S$ and $N$ the amplitudes of the sideband and the noisefloor in \si{dBm}, respectively.

\subsection{Measured device}
\begin{figure*}
	\centering
	\includegraphics[]{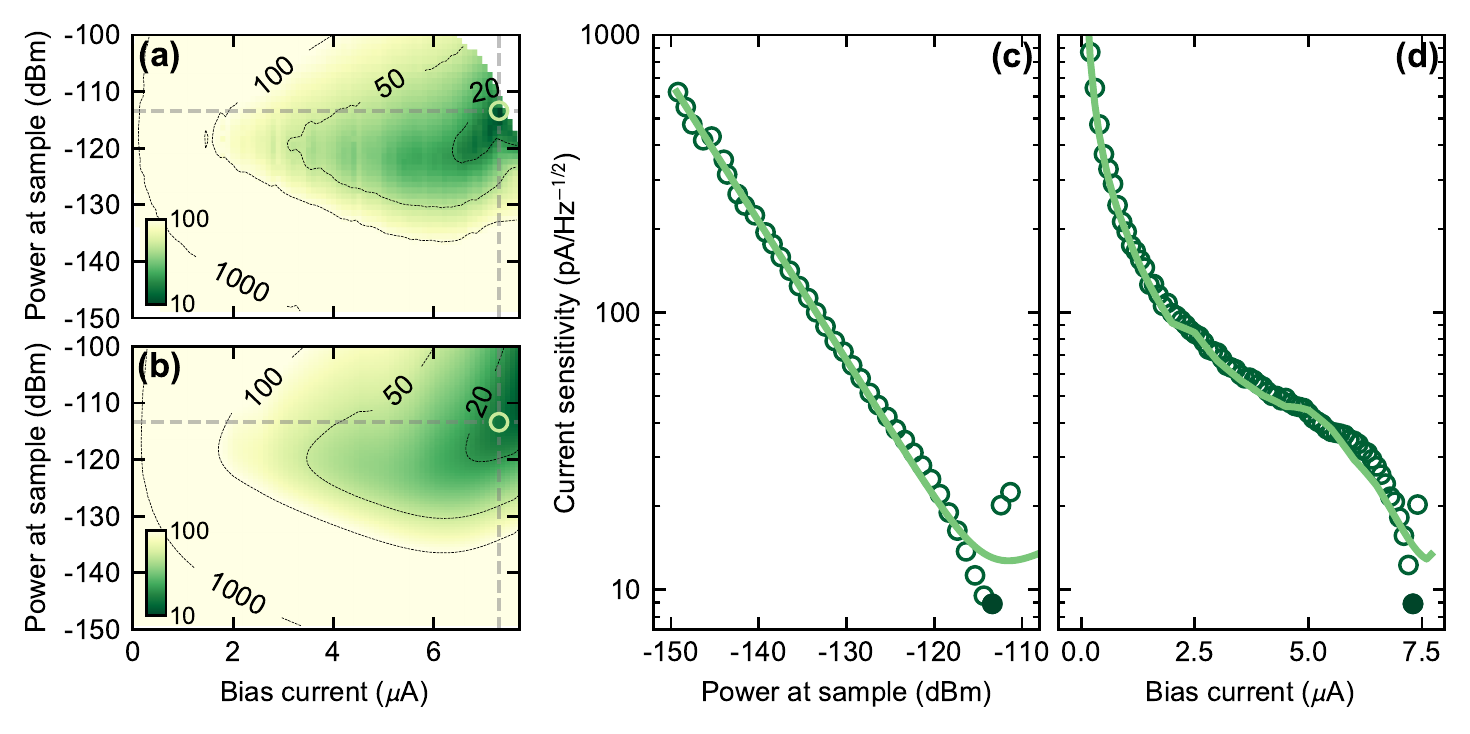}
	\caption{
		\textbf{Finding the best device sensitivity.}
		Current sensitivity in \si{\pico\ampere\per\sqrt{\hertz}} versus bias current and input power, as measured (a) and calculated (b).
		Dashed grey lines correspond to the linecuts in (c) and (d), circle marks the point of minimum measured sensitivity.
		Color scale is logarithmic from 10 to 1000, black lines mark contour lines of sensitivity values as labeled.
		(b) Sensitivity at \SI{7.3}{\micro\ampere} versus pump power (vertical line in (a,b)).
		We attribute discrepancies at high $P_\text{in}$ to differences in $\Delta$ between measurement and theory.
		(c) Sensitivity at $P_\text{in}=\SI{-113}{dBm}$ versus bias current (horizontal line in (a,b)).
		Circles: measured data, lines: model, full circle: minimum measured sensitivity.
		\label{fig:figure4}
	}
\end{figure*}

We analyze $\mathcal{S}_I$ for a large range of bias currents and drive powers.
The device sensitivities extracted via Eq.~\eqref{eq:sensitivity} are plotted in Figs.~\ref{fig:figure4}(a,b) for the measured and modeled data, respectively, showing good qualitative agreement.
Linecuts through the 2D measured and simulated data at the best measured value of $\mathcal{S}_I$ show good quantitative agreement between theoretical model and measurement, cf. Figs.~\ref{fig:figure4}(c,d).
For a fixed bias current, the current sensitivity drops exponentially as a function of input power, reaching a minimum value of $\mathcal{S}_I=\SI{8.9}{\pico\ampere\per\sqrt{\hertz}}$ at $I_b=\SI{7.3}{\micro\ampere}$ and $P_\text{in}=\SI{-113}{dBm}$.
Similarly, as a function of bias current and fixed input power, the current sensitivity drops rapidly over more than two orders of magnitude.
Our theoretical calculations deviate from the measured data for very large input powers and bias currents, for which the model predicts sensitivity values larger than observed.
This deviation might be due to minor differences in experimental and theoretical detuning:
If the the pump tone $\omega_0$ is slightly below the value of $\omega_0^\prime$ in the limit of $n_\text{ph}\rightarrow0$, the pump will initially be slightly red-detuned ($\Delta<0$) and move to blue-detuned ($\Delta>0$) as the resonance shifts downward due to the Kerr nonlinearity, instead of starting on-resonance and becoming only blue-detuned as we increase $P_\text{in}$.
Depending on the pump power at which $\Delta=0$, the theory curve will underestimate the sideband amplitude for $\Delta>0$, resulting in too large values of $\mathcal{S}_I$, as in Fig.~\ref{fig:figure4}(c) for $P_\text{in}\geq\SI{-120}{dBm}$.
As detailed in Sec.~\ref{sec:deviation_power} of the Supplemental Material~\cite{SeeSupplementalMaterial}, the model follows the measured data more closely for high pump powers assuming an initially red detuned drive.
This deviation is especially large for high bias currents because the anharmonicity grows with $I_b$.
Thus, the cavity resonance shifts stronger with pump power and the drive is more likely to have a smaller detuning than expected for high $P_\text{in}$.

\subsection{Limitations of present device and setup}\label{subsec:limitations}
Optimum sensitivity would be achieved for zero pump detuning, maximum pump power and biasing the device close to $I_c$, cf. Fig.\ref{fig:figure4}(a),(b).
In our experiment we were unable to operate the device in a stable regime for bias currents greater than $0.9 I_c$, after which the JJ occasionally switched to the normal state, destroying the RF resonance.
Additionally, we observed exponential increase of the internal loss rate for large bias currents.
These effects are presumably due to random phase diffusion across the junction and electrical interference in our setup, cf. Sec.~\ref{sec:lossrates} of the Supplemental Material~\cite{SeeSupplementalMaterial}.
Most notably, at elevated bias currents spurious sidebands at integer multiples of \SI{50}{\hertz} appear in the measured spectra, which are due to insufficient isolation between the DC and RF electronics.
Using the same approach as for the intended signal, we can quantify the current noise due to mains power to $\SI{168}{\pico\ampere}\approx I_\text{LF}/60$.
Improving the setup should allow us to move to even higher bias currents, gaining in $\mathcal{S}_I$.
In addition, the resonance frequency shift due to anharmonicity places an upper bound on the maximum input power.
In an optimized measurement, shifting the pump frequency with pump power in order to remain closer to resonance should allow us to gain more than \SI{10}{dB}, reaching a minimum of \SI{2.7}{\pico\ampere\per\sqrt{\hertz}}, cf. Sec~\ref{sec:drive_shift} of the Supplemental Material~\cite{SeeSupplementalMaterial}.

\subsection{Modeled optimized device}

\begin{figure}
	\centering
	\includegraphics[]{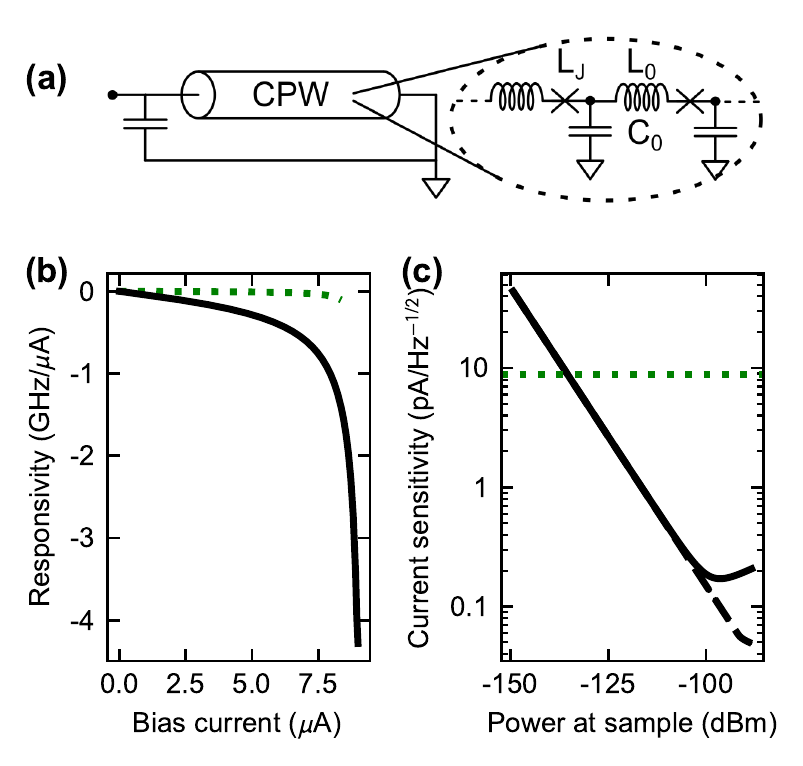}
	\caption{
		\textbf{Estimated sensitivities for optimized device design.}
		(a) Instead of a linear CPW shorted to ground by a nonlinear Josephson junction, the optimized device is a diluted JJ meta material with a CPW center conductor based on a Josephson junction array, directly shorted to ground.
		(b) Frequency responsivity $G_1$ for the optimized (solid line) and current device (dotted line).
		Due to the dominating Josephson inductance, the optimized device tunes further with bias current.
		(c) Predicted $\mathcal{S}_I$ for the optimized device.
		Dotted line indicates the minimum experimentally achieved sensitivity of \SI{8.9}{\pico\ampere\per\sqrt{\hertz}} with the present design.
		For the JJ array CPW, we predict sensitivities as low as \SI{170}{\femto\ampere\per\sqrt{\hertz}} (solid line).
		The sensitivity curves upwards for high pump powers due to the nonlinearity in the circuit.
		Choosing the pump frequency to be continuously close to resonance for high drive powers, the predicted sensitivity would drop down to \SI{50}{\femto\ampere\per\sqrt{\hertz}} (dashed line).
		Parametric amplification could reduce the sensitivity one order of magnitude further by reducing the contribution of the noise of the cryogenic HEMT amplifier in the readout noise of the cavity.
		\label{fig:figure5}
	}
\end{figure}

In order to improve $\mathcal{S}_I$, we propose a slightly changed circuit layout that follows naturally from the measured device and is immediately implementable:
Instead of a transmission line shorted to ground by a single JJ, we propose to incorporate the Josephson inductance into the transmission line itself, by means of a diluted JJ metamaterial~\cite{planatUnderstandingSaturationPower2019}.
The optimized design would then be a transmission line directly shorted to ground, with the CPW center conductor made up of a series of identical unit cells, each composed of a combination of linear and Josephson inductance ($L_0,L_J$) and a capacitance to ground ($C_0$), as depicted in Fig.~\ref{fig:figure5}(a).
Following the approach to circuit quantization presented in Ref.~\cite{gelyQuCATQuantumCircuit2019} and methods from Refs.~\cite{noscheseTridiagonalToeplitzMatrices2013,niggBlackBoxSuperconductingCircuit2012,vool_introductionquantum_2017}, we derive the resonance frequency of this CPW as
\begin{align}
	\omega_0(I_b)=\frac{\pi}{N\sqrt{C_0(L_J(I_b)+L_0)}} \ ,
\end{align}
in the limit of large $N$, as detailed in Sec.\ref{sec:optimized} of the Supplemental Material~\cite{SeeSupplementalMaterial}.
To maximize the responsivity $G_1$ of the device via maximizing the participation ratio $\eta_J=L_J/(L_0+L_J)$ per unit cell, we propose a CPW with center conductor and gap sizes $1/10$ of the current design and a reasonably short unit cell length of \SI{1}{\micro\meter}.
This would result in $L_0=\SI{842}{\femto\henry}$, $L_J=\SI{35.9}{\pico\henry}$ and $C_0=\SI{169}{\atto\farad}$ per unit cell, cf. Ref.~\cite{simonsCoplanarWaveguideCircuits2001} and Sec.~\ref{sec:optimized} of the Supplemental Material~\cite{SeeSupplementalMaterial}.
For an initial resonance frequency at $\omega_0=2\pi\times\SI{7.5}{\giga\hertz}$, the device would require approximately 845 unit cells, resulting in a total device length of \SI{845}{\micro\meter}, much more compact than our present layout.
Such an optimized device offers a significantly larger $G_1\approx\SI{4}{\giga\hertz\per\micro\ampere}$ with a relative frequency shift $\delta\omega_0/\omega_0\approx\SI{50}{\percent}$.
Additionally increasing the external coupling, e.g. by reducing the shunt capacitor to $1/4$ of its current size, this device would be able to achieve sensitivities as low as \SI{0.17}{\pico\ampere\per\sqrt{\hertz}}, a factor of 54 improvement to our presented design, as shown in Fig.~\ref{fig:figure5}(b).
We note that in an ideal experiment, the drive frequency should be tuned for increasing drive power in order to account for the Kerr-shift of the resonance to lower frequencies, thus minimizing $\Delta$ and maximizing $\alpha_0$.
Implementing this measurement scheme would allow us to achieve sensitivities down to \SI{50}{\femto\ampere\per\sqrt{\hertz}}.
Since this estimation does not take parametric amplification into account, we expect it to be an upper bound to the experimentally achievable $\mathcal{S}_I$:
Utilizing quantum-limited parametric amplification built into the device would allow us to gain approximately \SI{20}{dB}~\cite{stehlikFastChargeSensing2015,pogorzalekHystereticFluxResponse2017,planatUnderstandingSaturationPower2019}, providing noise levels down to \SI{5}{\femto\ampere\per\sqrt{\hertz}}.

\section{Conclusion}

We presented a Josephson parametric upconverter and demonstrated current sensitivities down to \SI{8.9}{\pico\ampere\per\sqrt{\hertz}} which makes our compatible with TES readout, and derived an analytical model that accurately reproduces the measured data and is immediately applicable to other device architectures.
We estimate that future devices using increased Josephson participation ratios, and using the intrinsic Kerr-nonlinearity for four-wave parametric amplification built into the detection cavity, should allow for an increase $\mathcal{S}_I\sim\SI{5}{\femto\ampere\per\sqrt{\hertz}}$, orders of magnitude better than state of the art KPUPs and limited by the fundamental quantum noise of the cavity.

\section*{Data availability}
All raw and processed data as well as supporting code for measurement libraries, data processing and figure generation is available in Zenodo~\cite{zenodo1}

\section*{Acknowledgements}
This project has received funding from the European Union Horizon 2020 research and innovation programme under grant agreement Nos. 681476 -- QOMD, 732894 -- HOT and 785219 -- GrapheneCore2.

\appendix

\section{Input-output formalism}\label{app:inputoutput}

Starting from Eq.~\eqref{eq:Duffing-EOM}, with the steady-state solution $\alpha_0$, the reflection coefficient is given by
\begin{align}
	S_{11}=-1-\sqrt{\kappa_\text{e}}\frac{\alpha_0}{S_\text{in}}=-1+\frac{2\kappa_\text{e}}{\kappa+2i\Delta}
\end{align}
where the second equality holds in the limit $\beta\rightarrow 0$ and which can be recognized as the usual reflection expression of circuit theory.

We now consider the case in which the cavity resonance frequency is a function of an additional current given by $I = I_b + \delta I$.
With the resonance frequency given by Eq.~\eqref{eq:omega_Taylor}, the new equation of motion reads
\begin{align}
	\begin{split}
	\dot{\alpha}=\left[ -i\left( \Delta -\sum_{m=1}^n G_m \delta I^m \right) - \frac{\kappa}{2} \right] \alpha \\
	-i\beta\abs{\alpha}^2\alpha +\sqrt{\kappa_\text{e}}S_\text{in}\ .
	\end{split}
\end{align}
With the Ansatz for the intracavity field $\alpha(t)=\alpha_0+\delta\alpha(t)$ and assuming $\abs{\alpha}^2 \approx \alpha_0^2$ we get
\begin{align}
	\begin{split}
		\delta\dot{\alpha}=\left[ -i\left( \Delta - \sum_{m=1}^n G_m \delta I^m \right) -\frac{\kappa}{2} \right] \delta\alpha \\
		+i \alpha_0 \sum_{m=1}^n G_m \delta I^m \ .
	\label{eq:delta-alpha-dot}
	\end{split}
\end{align}
Let the modulation in current be of the form
\begin{align}
	\delta I=I_\text{LF}\cos\Omega t = I_{-}e^{-i\Omega t} + I_{+}e^{+i\Omega t}
	\label{eq:deltaIs}
\end{align}
where $I_{-}=I_{+}=I_\text{LF}/2$.
Our Ansatz for $\delta\alpha$ is consequently
\begin{align}
	\delta\alpha=\sum_{m=1}^n a_{-m}e^{-mi\Omega t} + a_{+m}e^{+mi\Omega t}\ .
	\label{eq:deltaalpha}
\end{align}
Inserting Eqs.~\eqref{eq:deltaIs},\eqref{eq:deltaalpha} into Eq.~\eqref{eq:delta-alpha-dot}, we can group the terms by their frequency components and equalize each component individually in order to solve for the sideband coefficients $a_{\pm m}$.
Each sideband output field can then be calculated via
\begin{align}
	S_{\pm m} = \sqrt{\kappa_\text{e}} a_{\pm m}\ .
\end{align}
We arrive at a compact result for the first order sidebands appearing at $\omega_0\pm1\Omega$:
\begin{align}
	S_{\pm 1} = \frac{\sqrt{\kappa_\text{e}}\alpha_0 G_1 I_\text{LF}}{-i\kappa+2(\Delta\pm\Omega)}\ .
	\label{eq:input-output-first}
\end{align}

We calculated all $a_{\pm m}$ coefficients up to $m=3$ using \textit{Mathematica} v11.3.0.0 in the notebook \texttt{input-output formalism.nb}, which we subsequently converted to \textit{python3} code using the notebook \texttt{Export to Python.nb} located in Zenodo~\cite{zenodo1}.

\section{Calculating the steady-state solution}\label{sec:Duffing}
We can calculate $\alpha_0$ by solving Eq.~\eqref{eq:Duffing-EOM} for a large pump signal and treating the probe as a perturbation~\cite{castellanos-beltranDevelopmentJosephsonParametric2010}.
Thus, let us assume that the solution has the form $\alpha(t)=\alpha_0 \exp [i\omega_p t]$ and the input signal $S_\text{in}(t) = S_p(t) =S_{p0} \exp [i(\omega_p t+\phi)]$ is the pump signal.
Since we are only interested in the steady-state solution, let $S_{p0},\alpha_0 \in \mathbb{R}$.
Inserting this into Eq.~\eqref{eq:Duffing-EOM}, we get
\begin{align}
	\left(i\Delta+\frac{\kappa}{2}\right)\alpha_0 + i\beta\alpha_0^3=\sqrt{\kappa_\text{e}}S_{p0}e^{i\phi}
\end{align}
Multiplying this equation with its complex conjugate returns
\begin{align}
	\beta^2 \alpha_0^6 + 2\Delta\beta\alpha_0^4 + \left(\Delta^2+\frac{\kappa^2}{4}\right)\alpha_0^2 - \kappa_\text{e} S_{p0}^2 = 0\ .
	\label{eq:polynom}
\end{align}
While this third-order polynomial in $\alpha_0^2$ has multiple complex solutions, the ones relevant in our case are only real.
In the high-power regime, our resonator will exhibit bifurcation and Duffing behavior, meaning there will be three real valued solutions to $\alpha_0^2$:
The largest, median and smallest one corresponding to the high, middle and low amplitude branch, respectively.
For a given input field $S_{p0}$ and detuning $\Delta$, the (up to three) solutions of this equation can be found either numerically or analytically.
However, for the parameters used in our experiment, the solutions for $\alpha_0^2$ are identical because our drive remains outside of the bifurcation regime.
We can then use the corrected intracavity field for obtaining the sideband amplitudes by replacing the value of $\alpha_0$ for the linear oscillator in Eq.~\eqref{eq:input-output-first}.

Furthermore, taking the resonance frequency as the point where $\partial\alpha_0/\partial\omega=0$, we can compute the frequency shift the cavity experiences as a result of the driving power by differentiating Eq.~\eqref{eq:polynom} with respect to $\omega$ as
\begin{align}
	\omega_0^\prime = \omega_0 - \abs{\alpha_0}^2\beta\ .
	\label{eq:Duffing-shift}
\end{align}

\section{Higher order terms}\label{app:higher-orders}
\begin{figure}
	\centering
	\includegraphics[]{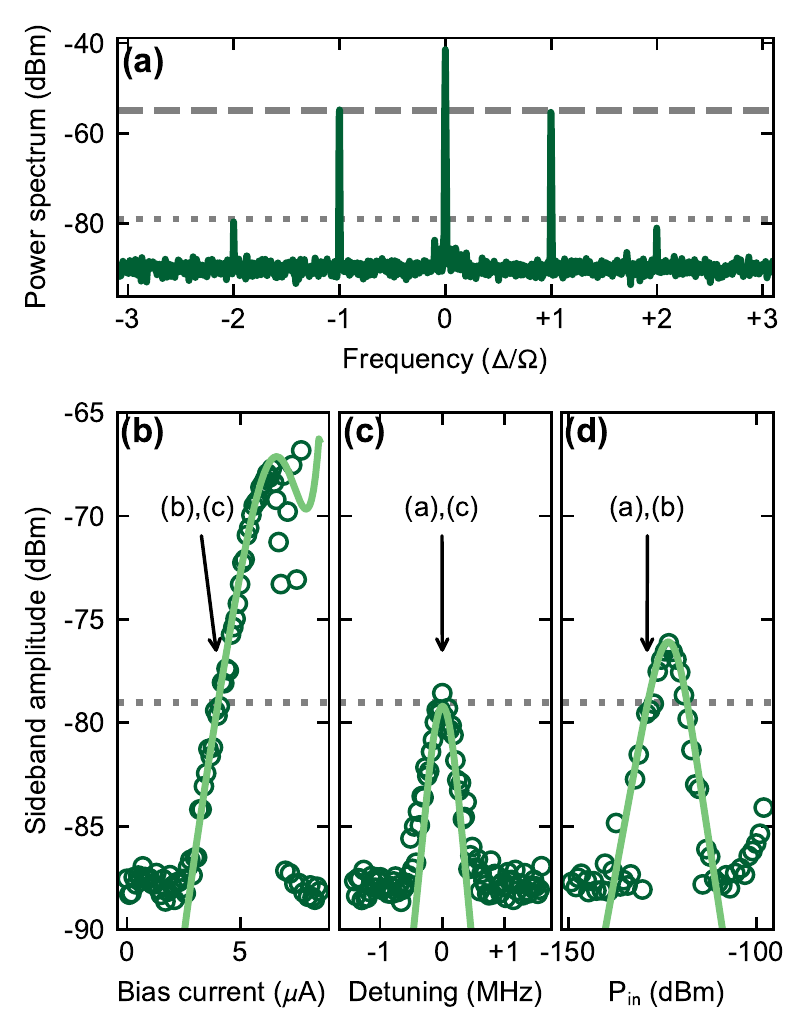}
	\caption{
		\textbf{Exploring the parameter space for the second order sideband amplitude.}
		(a) Sideband height for changing bias current setpoint at fixed input power and zero detuning for varying bias current.
		(b) Sideband height for changing drive detuning at fixed input power and bias current.
		(c) Sideband height for changing input power at fixed bias current and detuning.
		Circles: measured data, solid lines: calculated amplitude via input-ouput theory, dotted grey line: calculated sideband amplitude at $I_0=\SI{4}{\micro\ampere}$, $\Delta=0$ and $P_\text{in}=\SI{-129}{dBm}$.
		Arrows indicate the setpoints for the other respective panels.
		(d) The power spectrum at the output at $I_0=\SI{4}{\micro\ampere}$, $\Delta=0$ and $P_\text{in}=\SI{-129}{dBm}$ containing the input pump signal at $\omega_0$ and the first and second order sidebands due to mixing at $\omega_0 \pm \Omega$ and $\omega_0 \pm 2\Omega$.
		Dotted grey line corresponds to the one in panels (a-c), dashed grey line corresponds to the one in Fig.~\ref{fig:figure3}.
	}
	\label{fig:higher-order-peaks}
\end{figure}

Already for second order in $\delta I$, the prefactors are too complicated to write down in a short form, which is why we refer to the full analytical solutions in the \textit{Mathematica} notebook \texttt{input-output formalism.nb} located on Zenodo~\cite{zenodo1}.
We note that higher order corrections arising for terms in $\delta I^m$, have only negligible effects on the lower order forms.
For the analysis in the main text, we therefore only make use of the closed form for the first order terms, and for the second order peaks in Fig.~\ref{fig:higher-order-peaks}(b-d), only the second order terms were used.

We observe higher order sidebands over a wide range of operating points, with an exemplary spectrum exhibiting both first and second order peaks plotted in Fig.~\ref{fig:higher-order-peaks}(a).
Similar to Fig.~\ref{fig:figure3}(a), the second order sideband increases with DC bias current up to $I_b\approx 0.75 I_c$ where the amplitude is limited by the increasing internal loss rate, cf. Fig.~\ref{fig:higher-order-peaks}(b).
As depicted in Fig.~\ref{fig:higher-order-peaks}(c), finite drive detuning strongly suppresses the sideband amplitude similar to the first order peaks.
The power dependence, cf. Fig.~\ref{fig:higher-order-peaks}(c), also closely resembles the shape of the first order sideband, with maximum amplitude for high drive powers and subsequent decrease due to increasing drive detuning as a consequence of the downshift in resonance frequency due to the Kerr nonlinearity.

%


\pagebreak
\clearpage
\widetext

\setcounter{equation}{0}
\setcounter{figure}{0}
\setcounter{table}{0}
\setcounter{page}{1}
\setcounter{section}{0}

\renewcommand{\thepage}{S\arabic{page}}
\renewcommand{\thesection}{S\Roman{section}}
\renewcommand{\thetable}{S\Roman{table}}
\renewcommand{\thefigure}{S\arabic{figure}}
\renewcommand{\theequation}{S\arabic{equation}}

\textbf{\centering\large Supplementary Material:\\}
\textbf{\centering\large Current detection using a Josephson parametric upconverter\\}

\vspace{1em}

{\centering\noindent Felix E. Schmidt, Daniel Bothner, Ines C. Rodrigues, Mario F. Gely, Mark D. Jenkins, and Gary A. Steele$^{*}$\\}

{\centering\noindent\em Kavli Institute of NanoScience, Delft University of Technology, Lorentzweg 1, 2628 CJ, Delft, The Netherlands.\\}


\section{Measurement setup}\label{sec:measurement}

\subsection{Wiring configuration}

\begin{figure*}
	\centering
	\includegraphics[width=0.7\linewidth]{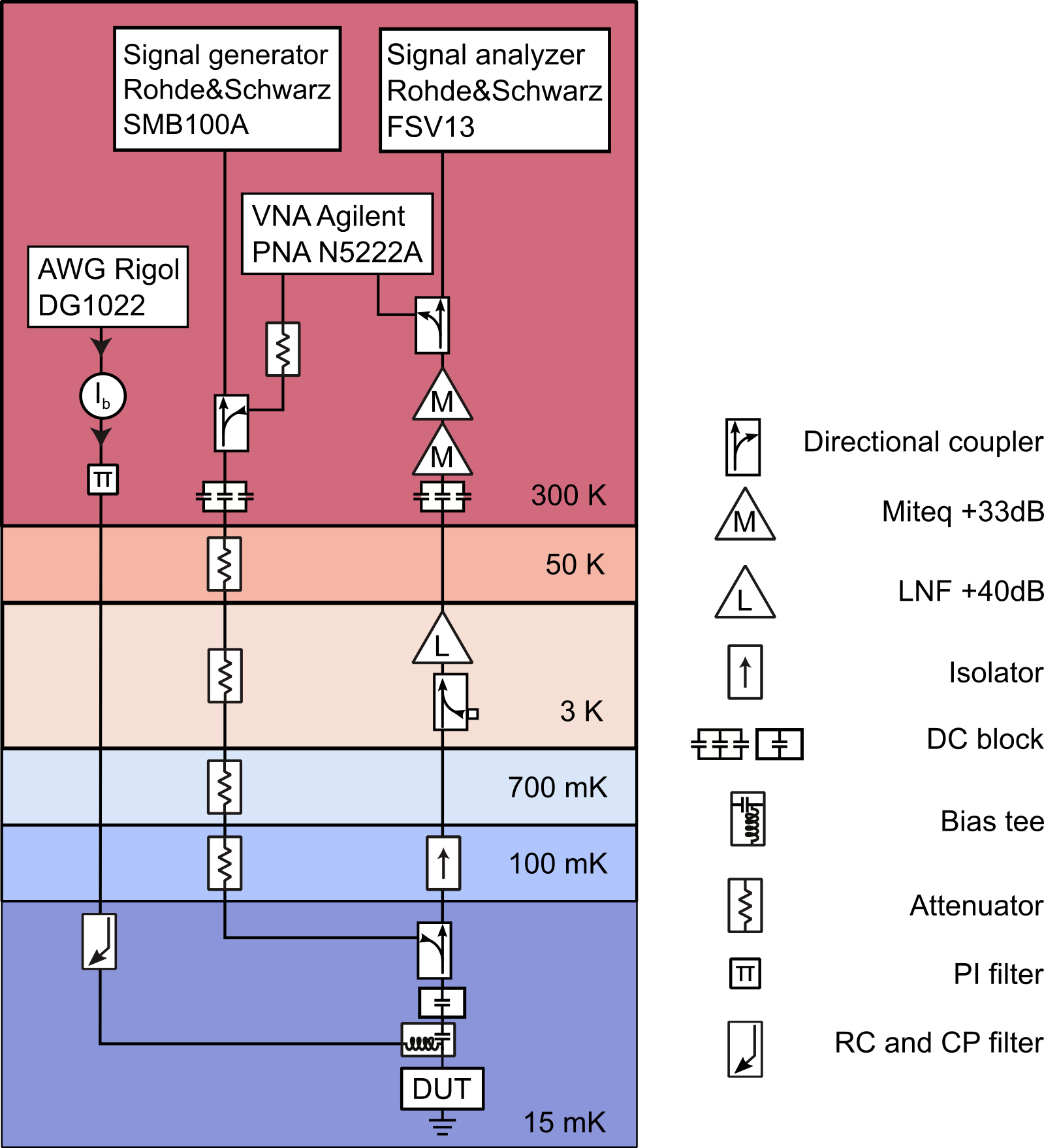}
	\caption{
		\textbf{Measurement setup used for all measurements.}
		The directional coupler at the \SI{3}{\kelvin} stage was terminated by a \SI{50}{\ohm} load at the coupled port.
		The bias current source is battery powered and voltage controlled, while all other equipment is mains powered.
		DC blocks at room temperature isolate both center and outer conductor, while the ones at \SI{15}{\milli\kelvin} only isolate the center conductor.
	}
	\label{fig:setup}
\end{figure*}

All measurements were taken with the device mounted to the millikelvin stage of a \textit{Bluefors BF 400-D} dilution refrigerator with a base temperature of approximately \SI{15}{\milli\kelvin}.
The measurement setup is sketched in Fig.~\ref{fig:setup}.
We use in-house built, low-noise battery powered electronics for DC biasing of the device.
For measurements involving current detection, we modulate the voltage controlled current source with an arbitrary waveform generator (AWG), model \textit{DG1022Z} from \textit{Rigol}.
Microwave reflection measurements of the cavity are done using a vector network analyzer (VNA) from \textit{Agilent}, model \textit{PNA N5222A}.
Signal generation and spectroscopy for current detection are done using signal generator \textit{SMB 100A} (SG) and analyzer \textit{FSV13} (SA), respectively, from \textit{Rohde \& Schwarz}.
The VNA and SG paths are merged using directional couplers.
Prior to the measurements on current detection, we calibrated the frequency dependent difference in attenuation between the signal paths VNA -- device under test (DUT) and SG -- DUT which we account for in all measurements and in the data analysis.

In order to minimize the influence of \SI{50}{\hertz} interference from mains powered equipment on our experiments, we place the DC electronics on an isolated rack and place all RF equipment on another one.
We observed significant signal deterioration for elevated bias currents if the DC and RF electronics shared the same ground.
For this reason, we placed additional DC blocks with separated inner and outer conductors on the RF input lines (PE8212 from \textit{Pasternack}).
Note that for the LF current modulation, we need to galvanically connect the AWG to our battery-powered voltage controlled current source, which in turn leads to a potential source of significant \SI{50}{\hertz} interference (see Sec.~\ref{sec:lossrates} for more elaborate discussion on this topic).

The DC lines are heavily filtered using $\pi$-filters inside the room-temperature electronics, and homemade copper powder and two-stage RC-filters on the baseplate of the dilution refridgerator.
The \SI{-3}{dB} cut-off for these filters is at around \SI{30}{\kilo\hertz}, well above the chosen modulation frequency of \SI{1}{\kilo\hertz}.

\subsection{Measurement protocol for current detection}
The measurements on current detection were performed using the following measurement scheme:
\begin{description}
	\item[1. Initialization and calibration] Turn off all outputs of RF instruments.
	Sweep the bias current back to zero and then to the next bias value.
	Set the VNA output power to low power and perform an $S_{11}$ measurement from \SI{7}{GHz} to \SI{8}{GHz}.
	From this measurement, determine the resonance frequency $f_0$ as the frequency at which $\abs{S_{11}}$ is minimum.
	\item[2. Current detection for fix pump power and detuning] Turn off the output of the VNA.
	Set the RF drive from the SG to low power and the drive frequency to $f_0$.
	Turn on the LF modulation ($I_\text{LF}=\SI{10}{\nano\ampere}$, $\Omega=2\pi\times\SI{1}{\kilo\hertz}$).
	Trigger the SAto perform one measurement.
	\item[3. Current detection for variable detuning] Keep the RF pump power at the same value and sweep the RF modulation frequency from $f_0-\SI{3}{\mega\hertz}$ to $f_0+\SI{3}{\mega\hertz}$ and for each detuning record the output signal using the SA.
	\item[4. Current detection for variable pump power] Set the RF frequency back to $f_0$.
	Sweep the RF pump power and for each pump power record the output signal using the SA.
	After each pump power measurement, reinitialize the bias current and find the resonance frequency again in order to reduce the number of "dead" cases in which the Josephson junction switched to the voltage state prematurely.
\end{description}

\subsection{Estimation of the attenuation and amplification chain}\label{sec:attenuation}

To estimate the attenuation chain, we use the thermal noise of our cryogenic high electron mobility transistor (HEMT) as a calibration source.
The noise power due to the effective noise temperature $T_{\rm e}\approx2{\,\rm K}$ of the HEMT as given by the manufacturer is
\begin{align}
P_{\rm N,in}^\prime=k_{\rm B}T_{\rm e}\Delta f \approx \SI{2.761e-23}{\watt\per\hertz} \times \SI{1}{\kilo\hertz} \approx \SI{2.761e-20}{\watt} \approx \SI{-165.6}{dBm}\ ,
\label{eq:HEMTnoise}
\end{align}
where $\Delta f=\SI{1}{\kilo\hertz}$ is the measurement bandwidth of our setup.
By averaging over a few $S_{11}$ traces taken with the VNA in an area unaffected by our DUT, i.e. off-resonant to the cavity and thus leaving the background unaltered in power, we extract an average signal and standard deviation which we use to define the signal-to-noise ratio at the VNA, $\text{SNR}_\text{VNA}=\SI{34.5}{\decibel}$ for a VNA output power of \SI{0}{dBm}.

In our setup, the added noise from the HEMT dominates over other noise sources, which we deduce from an increase in noise level when powering up the HEMT with the room temperature amplifiers already on.
Therefore, the SNR at the VNA is identical to the one at the HEMT output, and we deduce the power arriving at the HEMT input to be $P_{\rm in}=\SI{-131.3}{dBm}$. 
Between DUT and HEMT, the signal travels a certain distance of cabling and passes through additional microwave components, cf. Fig.~\ref{fig:setup}.
On the way, the signal will have been reduced by $X {\,\rm dB}$ due to the mentioned components, hence the power arriving at the HEMT will be $P_{\rm in}^\prime = 10^{-X/10} P_{\rm in}$, which results in an estimated attenuation of \SI{129.3}{dB} of our VNA input line, assuming $X=\SI{2}{dB}$ of cable loss between sample and HEMT.

We deduce the total gain of our amplification chain by calculating the average noise power measured with the SA, $P_\text{N,SA}=\SI{-97.5}{dBm}$ in a \SI{1}{\hertz} bandwidth, and substracting from it the HEMT noise power $P_{\rm N,in}^\prime=\SI{-195.6}{dBm}$ corresponding to the same bandwidth and the cable loss $X=\SI{2}{\decibel}$, resulting in a total gain of \SI{96.1}{\decibel} for the amplifier chain.

\section{Device fabrication}\label{sec:fabrication}

\begin{figure}
	\centering
	\includegraphics[width=.7\linewidth]{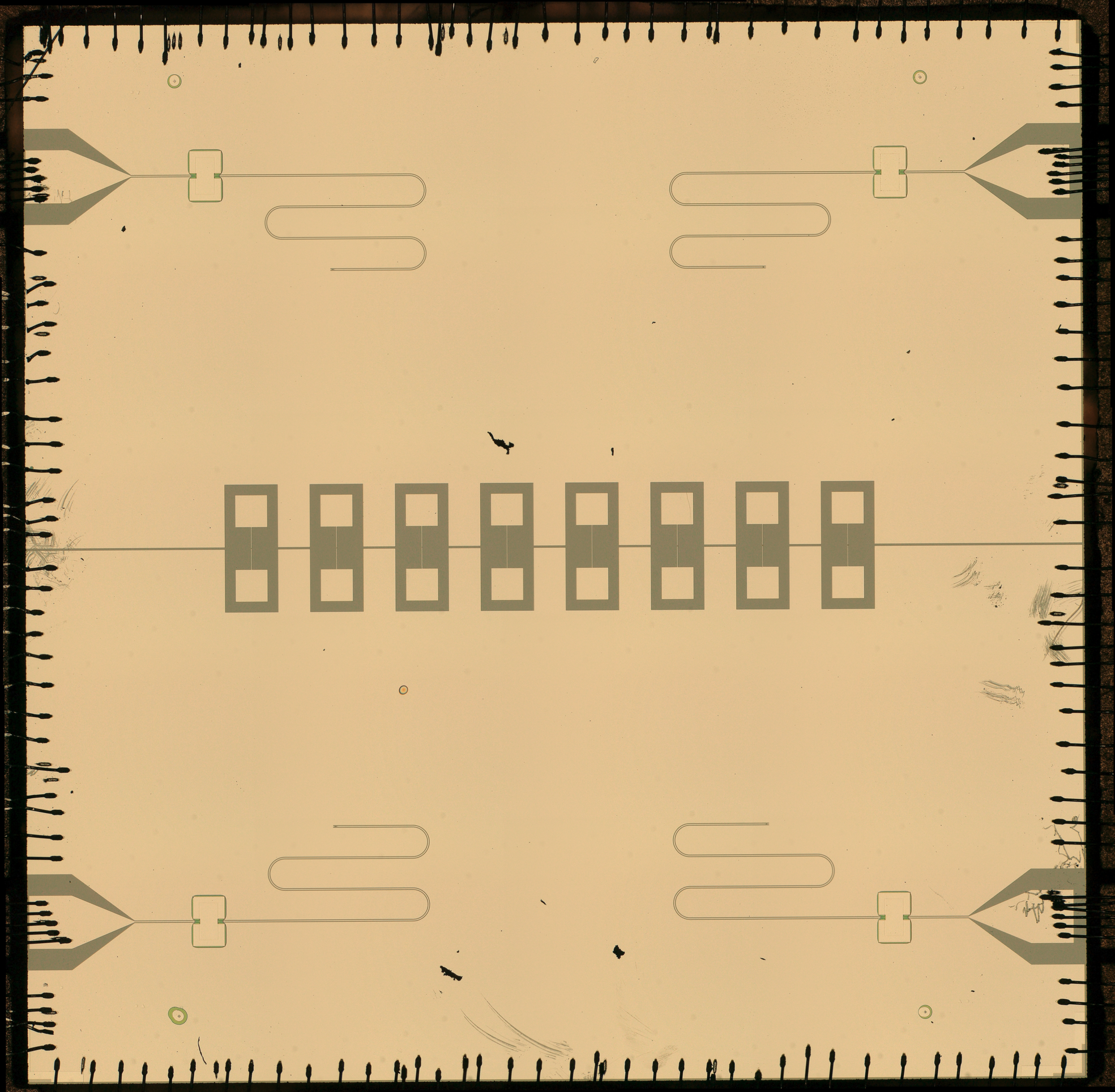}
	\caption{
		\textbf{The full chip.}
		Microscope image of the full chip, wirebonded to the PCB used for measurements.
		The chip hosts four different devices:
		The CPW with single JJ discussed in the main text (bottom left), a reference cavity with the same geometry shorted to ground (top left), and two devices shorted to ground by superconducting interference devices (SQUIDs) with loop sizes \SI{8x9}{\micro\meter} (top right) and \SI{5x5}{\micro\meter} (bottom right).
		Structures in the chip center are used for room-temperature DC tests.
		Chip size is \SI{10x10}{\milli\meter}.
	}
	\label{fig:fullchip}
\end{figure}

The device is fabricated in a four-step process in the \textit{Kavli Nanolab} cleanroom of TU Delft, using a combination of electron beam lithography (EBL, EBPG5000+ from \textit{Raith}), liftoff, sputtering (AC450 from \textit{Alliance}), PECVD (PlasmaPro 80 from \textit{Oxford Instruments}) and dry-etching (Fluorine reactive ion etcher from \textit{Leybold Hereaus}).
An optical micrograph of the fully packaged chip is shown in Fig.~\ref{fig:fullchip}.
The geometric device parameters are given in Table~\ref{tab:geometry}.
In the following we describe the fabrication step by step.
\begin{description}
	\item[Substrate] We use a double-side polished high-resistivity ($>\SI{6}{\kilo\ohm\centi\meter}$, light P/Boron doping, \SI{550}{\micro\meter} thickness) \SI{4}{inch} silicon wafer from \textit{IWS} as substrate for our device.
	The wafer is covered in positive electron beam resist (AR-P 6200.13, approximate thickness \SI{550}{\nano\meter}) and exposed to define the pattern for alignment and dicing markers.
	We sputter-deposit \SI{50}{\nano\meter} of Molybdenum-Rhenium (MoRe, RF magnetron sputtering in argon atmosphere from a \SI{60}{\percent} Mo-\SI{40}{\percent} Re target) and lift off the resist-protected areas using an anisole bath and strong ultrasonication, followed by multiple acetone and isopropanol baths.
	We subsequently cover the wafer with photoresist (HPR 504, \SI{1.2}{\micro\meter} thick) and dice it into \SI{14x14}{\milli\meter} chips for easier handling during fabrication.
	\item[Base layer] We pattern the Josephson junction together with the base layer and ground planes in a single lift-off step using AR-P 6200.09 (\SI{200}{\nano\meter}) and \SI{20}{\nano\meter} of sputtered Aluminum-Silicon (AlSi, reactive DC magnetron sputtering in argon atmosphere from a \SI{99}{\percent} Al-\SI{1}{\percent} Si target).
	Lift-off is done by placing the chip in the bottom of a beaker with room-temperature anisole and strong ultrasonication for a few minutes.
	\item[Dielectric layer] For the shunt dielectric layer, we deposit \SI{140}{\nano\meter} amorphous silicon at \SI{90}{\celsius} using PECVD.
	Patterning is done with EBL of a double-layer resist (PMMA 950K A4 and AR-N 7700.18) and reactive ion etching in a \ce{SF6 + He} atmosphere.
	The resist layers are in-situ removed using \ce{O2} plasma.
	\item[Top shunt plate] The top plate of the shunt capacitor is fabricated with an additional lift-off step using the same resist as for the alignment markers, and sputtering \SI{100}{\nano\meter} \ce{AlSi}.
	\item[Packaging] To fit our printed circuit board (PCB), the chip is again covered in photoresist and trimmed down to \SI{10x10}{\milli\meter}.
	After washing off the photoresist in a series of acetone and isopropanole baths, the chip is glued to our copper sample holder, to which the PCB is mounted, using cryogenic GE varnish.
	Electrical connections to the device are made using wedge-bonding on a Westbond wirebonder with aluminum wire bonds.
	We place a small copper lid on the chip to protect it from dirt and to suppress box modes of a bigger copper lid screwed onto the copper base, which accomodates the SMA connectors.
\end{description}

\begin{table}
	\caption{Geometric device parameters\label{tab:geometry}}
	\begin{tabular}{ccc}
		\hline \hline
		Symbol       & Description                                                                 & Value                            \\
		\hline
		$s$          & CPW center conductor                                                        & \SI{10}{\micro\meter}            \\
		$w$          & CPW gaps to ground                                                          & \SI{6}{\micro\meter}             \\
		$t$          & Base layer thickness                                                        & \SI{20}{\nano\meter}             \\
		$L_g^\prime$ & Geometric inductance per length \cite{simonsCoplanarWaveguideCircuits2001}  & \SI{424}{\nano\henry\per\meter}  \\
		$C_0^\prime$ & Geometric capacitance per length \cite{simonsCoplanarWaveguideCircuits2001} & \SI{169}{\pico\farad\per\meter}  \\
		$l$          & CPW length, from end of shunt to JJ                                         & \SI{6382}{\micro\meter}          \\
		$A_s$        & Shunt capacitor area                                                        & \SI{57800}{\micro\meter\squared} \\
		$t_d$        & Dielectric layer thickness                                                  & \SI{140}{\nano\meter}            \\
		\hline\hline
	\end{tabular}
\end{table}

\section{General device parameters}\label{sec:general}

\subsection{Reflection coefficient}\label{sec:general-S11}
The reflection coefficient of a transmission line with a shunt capacitor to ground on the input side and shorted to ground on the far end is given by
\begin{align}
S_{11}(\omega,\kappa_\text{i},\kappa_\text{e}) = \frac{\kappa_\text{e}-\kappa_\text{i}-2i\Delta}{\kappa_\text{e}+\kappa_\text{i}+2i\Delta} = -1+\frac{2\kappa_\text{e}}{\kappa+2i\Delta}
\label{eq:S11simple}
\end{align}
with the detuning from resonance, $\Delta=\omega-\omega_0$ and the internal, external and total loss rates $\kappa_i$, $\kappa_e$ and $\kappa=\kappa_\text{e}+\kappa_\text{i}$.

\begin{figure}
	\centering
	\includegraphics[]{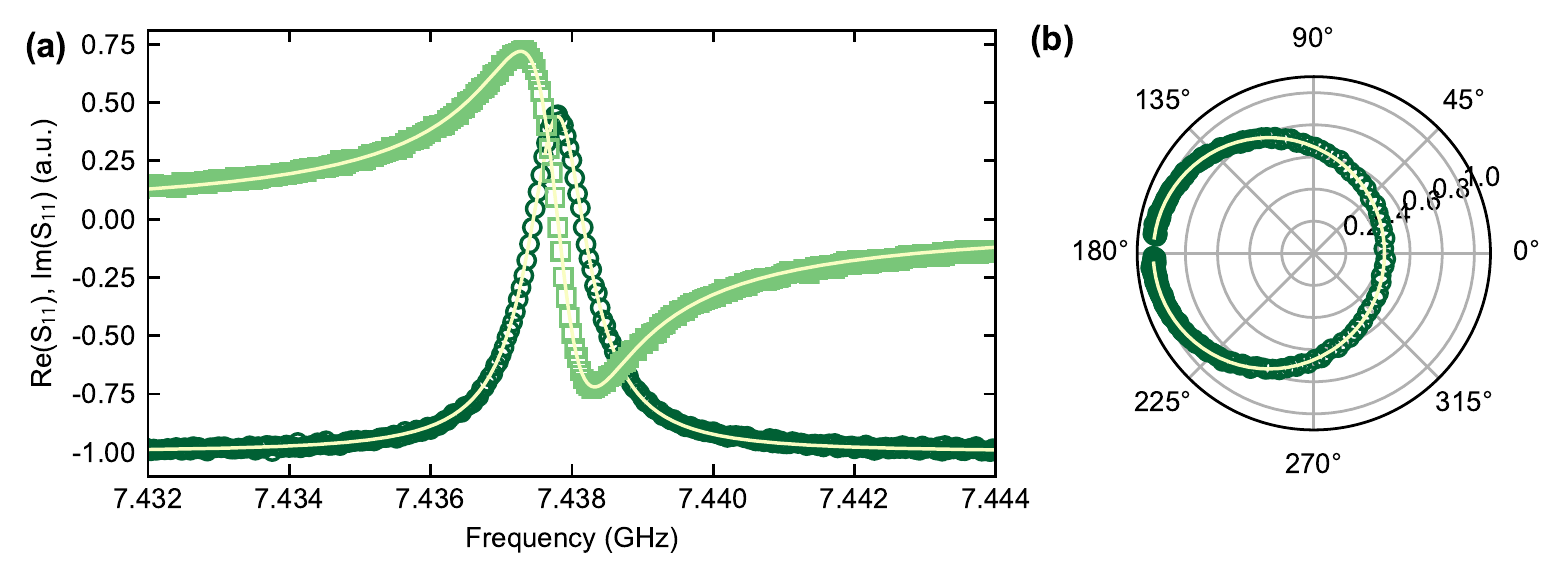}
	\caption{
		\textbf{Data and fit of the reflection coefficient.}
		(a) Real ($\bigcirc$) and imaginary ($\square$) value of the reflection coefficient $S_{11}$ versus frequency.
		(b) Polar plot of the absolute value of $S_{11}$ versus phase.
		Markers: data, solid lines: fits according to Eq.~\eqref{eq:S11simple}.
		The complex background and rotation are removed in both panels.
		\label{fig:s11fitabs}
	}
\end{figure}

The real response function is however distorted by the complex microwave background which arises due to impedance mismatches in our measurement setup.
For this reason, we model the measured $S_{11}$ spectra using the above model for an ideal device multiplied by a complex microwave background and a rotation in the complex plane:
\begin{align}
S_{11}^\prime(\omega,\kappa_\text{i},\kappa_\text{e},\theta) & = \left( a+b\omega+c\omega^2 \right) e^{i(a^\prime + b^\prime \omega)}  \left\{e^{i\theta}\left[S_{11}(\omega,\kappa_\text{i},\kappa_\text{e})+1\right]-1\right\}
\label{eq:S11full}
\end{align}
Our fitting algorithm first detects the resonance as frequency corresponding to the maximum phase derivative and fits the background signal by removing a certain window around the resonance frequency.
In a second step, it fits the modified model to the full data set keeping the background parameters fixed, and finally refits all background and model parameters once more starting from the previously fitted values.
A result of this fitting procedure, with background removed, is shown in Fig.~\ref{fig:s11fitabs}.

\subsection{Kinetic inductance estimation}\label{sec:Lk}

\begin{figure}
	\centering
	\includegraphics[]{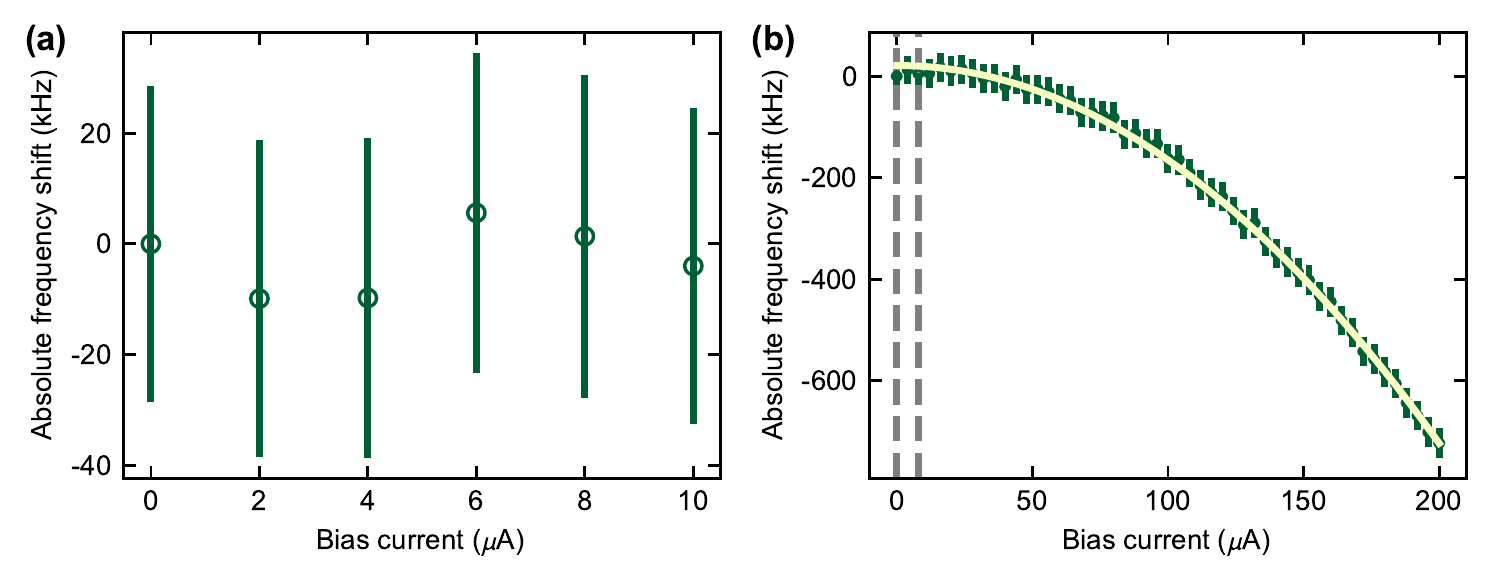}
	\caption{
		\textbf{Frequency shift of the reference CPW shorted to ground.}
		(a) Within the bias range for the device in the main text, $I_\text{0,max}=\SI{8}{\micro\ampere}$, the fluctuations in resonance frequency of the reference device are not due to kinetic inductance but remain within the range of our fit errors for the resonance frequency.
		(b) Biasing the reference resonator up to \SI{200}{\micro\ampere} allows us to extract the characteristic current for the kinetic inductance of the superconductor, $I_*\approx\SI{7.35}{\milli\ampere}$.
		Markers: data with error bars, solid line: fit to Eq.~\eqref{eq:kinetic-tuning}, dashed lines: range of applied bias currents for JPUP experiments in main text.
	}
	\label{fig:reference-current}
\end{figure}

Our AlSi films have a significant kinetic inductance contribution due to their small thickness of only \SI{20}{nm}.
We estimate the kinetic inductance fraction by performing a finite element electromagnetic simulation using \textit{Sonnet} v16.56 (Sonnet Software Inc., 2018) of the reference device (shorted to ground on the same chip, thus excluding the Josephson inductance) which results in an expected resonance due to only geometry at $\omega_{g}=2\pi\times\SI{8.83}{\giga\hertz}$.
We compare this value with the measured value of $\omega_{k}=2\pi\times\SI{7.56}{\giga\hertz}$.
The kinetic inductance fraction is given by~\cite{gaoExperimentalStudyKinetic2006}
\begin{align}
\eta_k = \frac{L_k^\prime}{L_k^\prime+L_g^\prime}=1-\left(\frac{\omega_k}{\omega_g}\right)^2
\end{align}
which has a value of $0.267$ in our device, hence $L_k^\prime=\SI{154}{\nano\henry\per\meter}$.
Kinetic inductance also increases as a function of DC bias current via $L_k(I)=L_k(0)\left[1+(I/I_*)^2\right]$ with the characteristic current $I_*$ \cite{annunziataTunableSuperconductingNanoinductors2010}.
The resulting downshift of the resonance frequency can be described by
\begin{align}
\omega_0(I_b) = \frac{\omega_0(0)}{\sqrt{1+\eta_kI_b^2/I_*^{2}}} \ .
\label{eq:kinetic-tuning}
\end{align}
We emphasize however that in the JJ device, the sheet kinetic inductance is not the relevant tuning parameter:
Biasing the reference device up to \SI{10}{\micro\ampere} does not show a trend; instead the fluctuations in $f_0$ remain within the fitting errors, cf.  Fig.~\ref{fig:reference-current}(a).
Only when applying bias currents up to \SI{200}{\micro\ampere}, well beyond the values used for the measurements presented in the main text, does the resonance frequency of the reference device shift to lower frequencies, as shown in Fig.~\ref{fig:reference-current}(b).
We fit the data using Eq.~\eqref{eq:kinetic-tuning}, extracting $I_*\approx\SI{7.35}{\milli\ampere}$.
This supports our claim that we can exclude kinetic inductance as an additional source of tuning the resonance frequency via applied bias currents and instead identify the Josephson inductance as the relevant tuning parameter.

\subsection{Resonance frequency versus current due to Josephson inductance}\label{sec:resfit}

Adapting the calculation for a Josephson terminated transmission line cavity given in Ref.~\cite{pogorzalekHystereticFluxResponse2017} to our device, we find for the current dependence of the cavity resonance frequency
\begin{align}
\omega_0(I_b)=\frac{\omega_{\lambda/2}}{1 + L_J(I_b,I_c)/L_r}\ ,
\end{align}
with $L_r$ the total device inductance, $\omega_{\lambda/2}$ the resonance frequency without the JJ and $L_J$ the Josephson inductance given in Eq.~\eqref{eq:Lj-of-I} of the main text.
We use this model to fit the measured resonance frequencies and extract $\omega_{\lambda/2}=2\pi\times\SI{7.515}{\giga\hertz}$, $L_r=\SI{3.458}{\nano\henry}$ and $I_c=\SI{9.176}{\micro\ampere}$.
From the device geometry and kinetic inductance estimation, we expect $L_r=(L_g^\prime+L_k^\prime)l=\SI{3.689}{\nano\henry}$, which is in acceptable agreement with the fit value.

\section{On the hysteresis of switching currents in our DC measurements}\label{sec:hysteresis}
Accodring to a simple RCSJ model, hysteresis in Josephson junctions should occur only for JJ quality factors $Q=R\sqrt{2eI_c C_J/\hbar} > 1$, with the junction capacitance $C_J$~\cite{tinkhamIntroductionSuperconductivity1996}.
We perform finite element simulations with \textit{Sonnet} v16.56 (Sonnet Software Inc., 2018) to estimate the stray capacitance of the metal leads in direct vicinity of the JJ (two \SI{1x1}{\micro\meter} pads separated by \SI{200}{\nano\meter}) to be $C_J=\SI{5.2}{\femto\farad}$.
Together with $R=\SI{108}{\ohm}$ and $I_c=\SI{9.176}{\micro\ampere}$, the junction would have $Q\approx1.3$.
For $Q \gtrsim 1$, the ratio between retrapping and critical current can be approximated as $I_r/I_c=4/\pi Q$.
Hence, in order to satisfy our measured values, $Q=4/\pi\times I_c/ I_r\approx1.91$, which would be reached for $C_J\approx\SI{13}{\femto\farad}$.
Likely, the geometric capacitance of the CPW and the surrounding ground planes significantly contributes and dominates the circuit capacitance:
Already including a \SI{50}{\micro\meter} portion of the CPW increases $C_J$ to \SI{14}{\femto\farad}, satisfying this requirement.
Additionally, we note that local heating in the junction area can also play a significant role, reducing $I_r$ further~\cite{skocpolSelfHeatingHotspots1974,hazraHysteresisSuperconductingShort2010,kumarReversibilitySuperconductingNb2015}.

\section{On the increased loss rates for increased bias current}\label{sec:lossrates}

\begin{figure}
	\centering
	\includegraphics[]{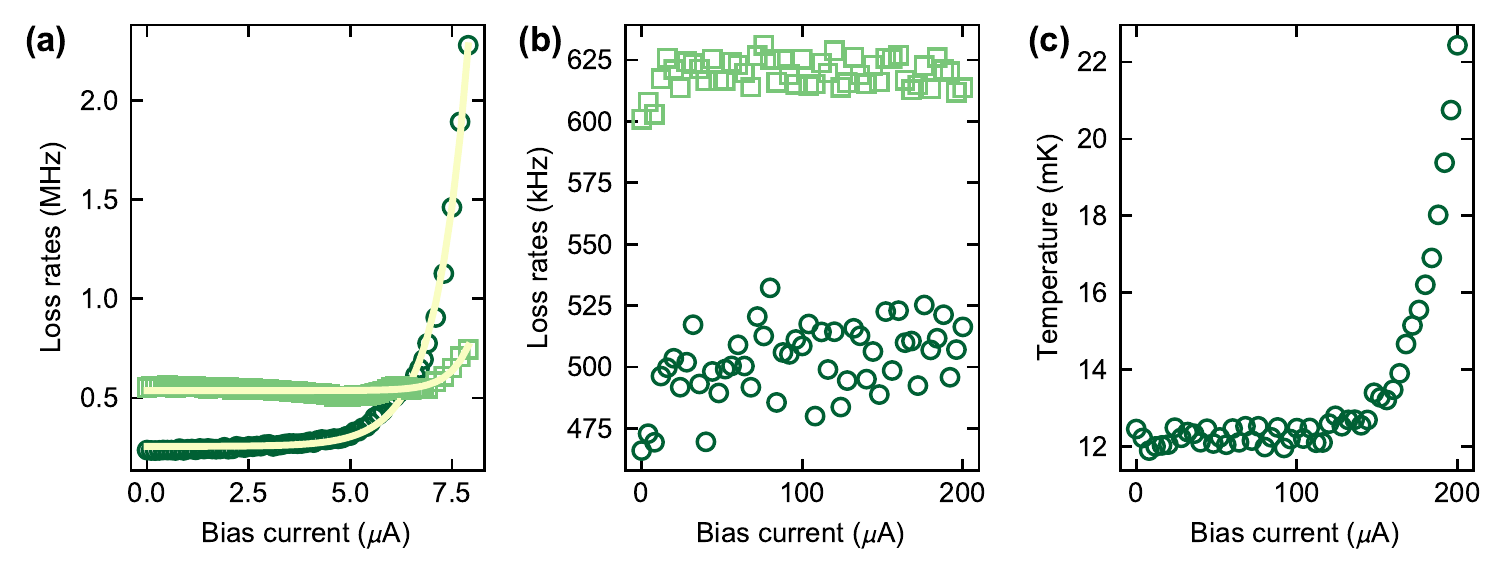}
	\caption{
		\textbf{Bias current dependent cavity loss rates.}
		(a) Internal ($\bigcirc$) and external ($\square$) loss rates versus bias current of the JJ terminated device. The measured values can be described by the exponential model (lines) from Eq.~\eqref{eq:lossrates}.
		(b) Internal ($\bigcirc$) and external ($\square$) loss rates of the reference device.
		We do not observe an increase in loss rate up to \SI{200}{\micro\ampere}.
		(c) Base temperature during measurement of (b).
		As we exceed the fridge cooling power, the mixing chamber plate heats up, but without any clear effect on $\kappa_\text{i}$ and $\kappa_\text{e}$.
	}
	\label{fig:lossratesvscurrent}
\end{figure}

\begin{figure*}
	\centering
	\includegraphics[]{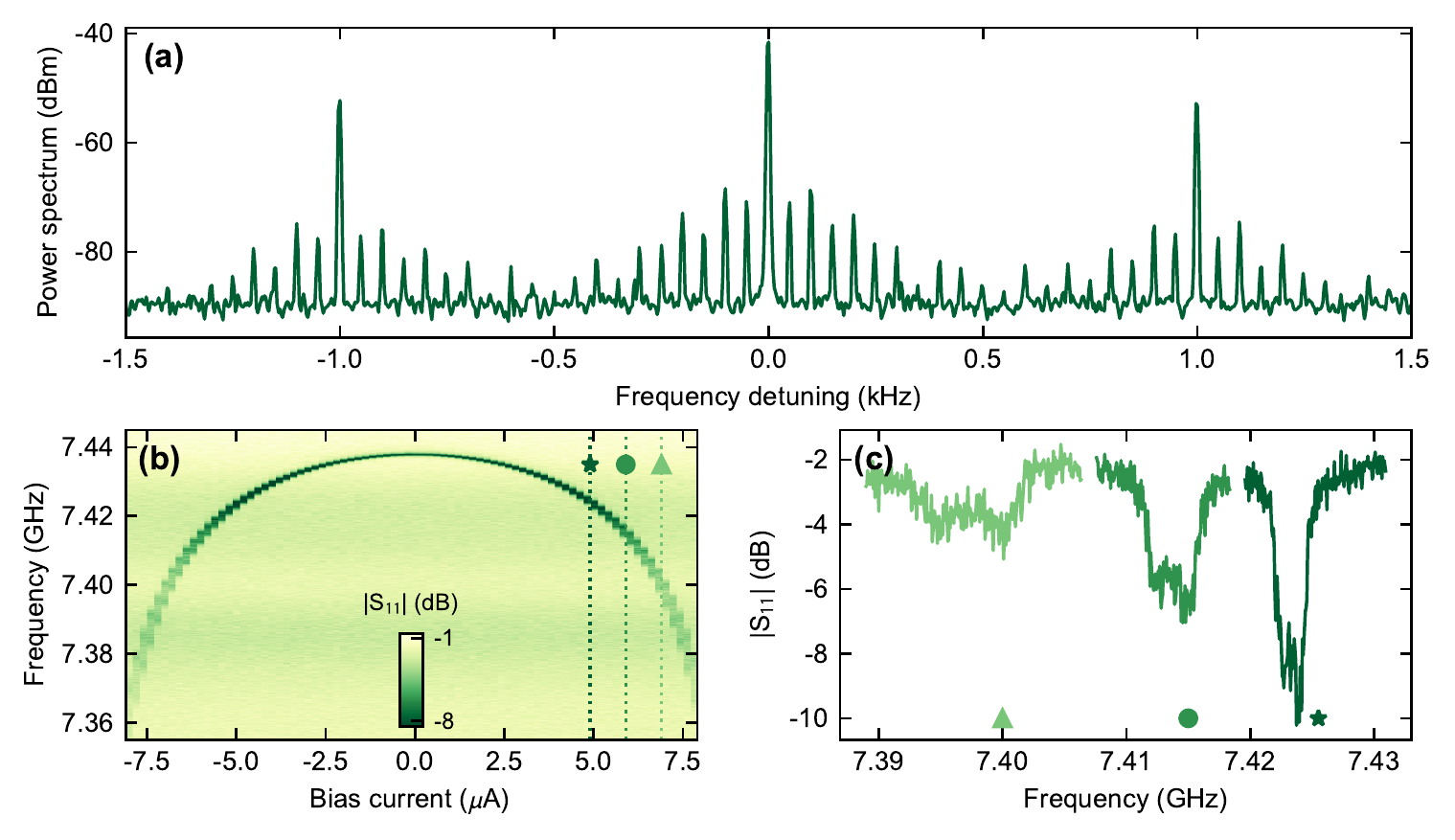}
	\caption{
		\textbf{Mains power interference.}
		(a) Current-mixing output spectrum taken at \SI{7.1}{\micro\ampere} bias current for the device in the main text.
		Peaks at zero frequency and $\pm\SI{1}{\kilo\hertz}$ correspond to the cavity-resonant continuous wave pump tone and the mixing sidebands, respectively.
		Even with mains powered equipment separated from battery electronics, spurious peaks appear at integer multiples of \SI{50}{\hertz}, corresponding to current signals of approximately \SI{170}{\pico\ampere}.
		(b) $\abs{S_{11}}$ measurements of the same device in the same cooldown, but with \SI{50}{\hertz} interference coupling to the DC electronics via a common ground.
		(c) Linecuts through (b) at various bias currents as indicated in (b).
		For high bias currents, the resonance splits into two dips, clearly hinting at significant current noise, which we estimate to be \SIrange{150}{250}{\nano\ampere}.
	}
	\label{fig:50hzinterference}
\end{figure*}

We observed an increase in the internal and external loss rates of the JJ terminated device for increased bias current, as can be seen in Fig.~\ref{fig:lossratesvscurrent}(a).
We can fit the loss rates quite accurately with a phenomenological exponential model of the form
\begin{align}
\kappa(I_b) = \kappa_0 + \kappa_1 \exp\left[\frac{I_b}{I_{\sim}}\right]\ .
\label{eq:lossrates}
\end{align}
The extracted parameters for the device presented in the main text are given in Tab.~\ref{tab:lossrates}.
While $\kappa_\text{e}$ should not directly depend on bias current, we do observe a slight increase, possibly due to changes in the impedance, linewidth broadening due to the increased $\partial\omega_0/\partial I_b$ or shifting the cavity through cable resonances.
Regarding the increase in internal loss rate, we identified the following mechanisms as most likely:
\begin{description}
	\item[Electrical interference] While we physically disconnect all mains-powered equipment from our battery powered DC electronics, and placed DC blocks for both inner and outer conductors on the RF inputs to the fridge, we still notice a significant amount of \SI{50}{\hertz} interference on the measured spectra for high bias currents, cf. Fig.~\ref{fig:50hzinterference}(a).
	We calculate the magnitude of the spurious signals to be approximately \SI{170}{\pico\ampere}, or \SI{1.7}{\percent} of $I_\text{LF}$.
	We assume that this interference is always present but only has noticeable effects for large $\partial\omega_0/\partial I_b$.
	Since our RF spectroscopy measurement takes more than \SI{1}{\second}, the measured linewidth is effectively broadened by the moving cavity, induced by small-scale \SI{50}{\hertz} modulations.
	Note that the situation is significantly worse for an unoptimized setup (noise coupling into the DC electronics via ground):
	In this case, the cavity spectrum is extremely broadened and starts to resemble two dips for high $I_0$, severely limiting the tuning range (c.f. Fig.~\ref{fig:50hzinterference}(b) without, and Fig.~\ref{fig:figure1}(f) of the main text with isolation of the battery electronics from mains powered equipment).
	We estimate the interfering current signal for this case to be \SIrange{150}{250}{\nano\ampere}, significantly limiting any device operation in either DC or RF.
	The measurement setup could be further improved by choosing a LF modulation frequency which is not a higher order multiple of \SI{50}{\hertz}, such as \SI{1111}{\hertz}, instead of \SI{1000}{\hertz}.
	We note that there might be other frequencies at which interfering signals couple into our device.
	Electrical noise only cannot explain the observed increased loss rates because the latter are not simply proportional to $\partial\omega_0/\partial I_b$.
	\item[Phase diffusion] The shorted reference device exhibits constant $\kappa_\text{i}$ and $\kappa_\text{e}$ upon DC bias up to \SI{200}{\micro\ampere}, at which point the mixing chamber starts to heat as we surpass the cooling power  of \SI{14}{\micro\watt} for the \textit{LD400 Bluefors} since the power dissipated in our \SI{2.7}{\kilo\ohm} low-pass filters reaches up to \SI{108}{\micro\watt}.
	In contrast, the temperature did not increase when measuring the JJ device.
	We therefore rule out quasiparticles (due to radiation or thermal excitations, \cite{tinkhamIntroductionSuperconductivity1996})in the CPW as a significantly contributing loss mechanism.
	Instead, phase diffusion across the JJ can indeed play a significant role as the bias current approaches $I_c$.
	Applying a DC current bias tilts the Josephson energy potential, enhancing the chance of phase-slip and quantum tunneling events, which can lead to dissipation even without switching to the normal state, as long as the phase particle is able to settle in the next washboard-potential minimum \cite{kiviojaWeakCouplingJosephson2005}.
\end{description}

\begin{table}
	\caption{Extracted loss rate parameters for the main text device.\label{tab:lossrates}}
	\begin{tabular}{cccc}
		\hline \hline
		Loss channel         & $\kappa_0/2\pi$ (\si{\kilo\hertz}) & $\kappa_1/2\pi$ (\si{\hertz}) & $I_{\sim}$ (\si{\nano\ampere}) \\
		\hline
		Internal & $256.16 \pm 3.74$             & $78.41 \pm 11.86$      & $777.89 \pm 11.96$             \\
		External & $537.41 \pm 2.11$             & $0.021 \pm 0.035$      & $487.43 \pm 52.40	$             \\
		\hline \hline
	\end{tabular}
\end{table}

\section{Difference between first order red and blue sidebands}\label{sec:bluered}
\begin{figure}
	\centering
	\includegraphics[]{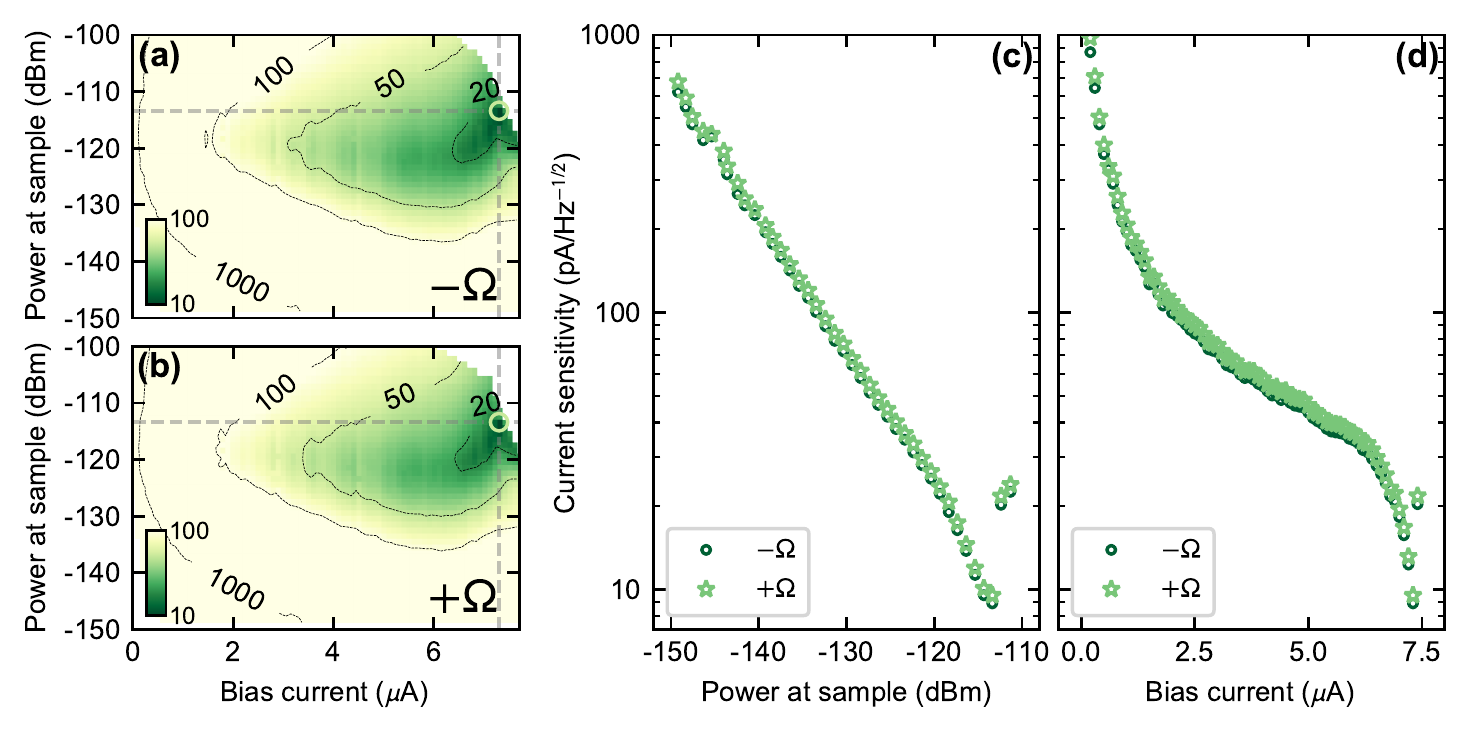}
	\caption{
		\textbf{Comparison of blue and red sideband.}
		Current sensitivity in \si{\pico\ampere\per\sqrt{\hertz}} versus bias current and input power, as measured for $-\Omega$ (a, cf. Fig.~\ref{fig:figure4} of the main text) and $+\Omega$ (b).
		Dashed grey lines correspond to the linecuts in (c) and (d), circle marks the point of minimum measured sensitivity.
		(b) Sensitivity at \SI{7.3}{\micro\ampere} versus pump power (vertical line in (a,b)).
		(c) Sensitivity at \SI{-113}{dBm} versus bias current (horizontal line in (a,b)).
		$\circ$: $-\Omega$ sideband, $\star$: $+\Omega$ sideband.
		Minimum sensitivities are \SI{8.90}{\pico\ampere\per\sqrt{\hertz}} and \SI{9.52}{\pico\ampere\per\sqrt{\hertz}}, respectively.
		Within the traced out parameter space, we observed no systematic differences between red and blue first order sidebands.
	}
	\label{fig:plusminus}
\end{figure}

As derived in the main text, to first order in $\delta I$, the coefficients describing the sideband amplitudes are
\begin{align}
a_{\pm 1} & = \frac{\alpha_0 G_1 I_\text{LF}}{-i\kappa+2(\Delta\pm\Omega)} \ .
\label{eq:input-output-first}
\end{align}
Without changing the device, our peak height will increase if we modulate the current stronger ($I_\text{LF}\uparrow$) or slower ($\Omega\downarrow$), and by pumping harder ($S_\text{in}\uparrow$).
With the same setup, increasing the current responsivity ($G_1\uparrow$) would likewise enhance the peak height.
Moreover, for small LF modulations $\Omega\ll\kappa$, the peaks for red and blue sidebands should be equal in amplitude.
In fact, the absolute difference between the two sidebands scales with
\begin{align}
\abs{a_{-1}}^2-\abs{a_{+1}}^2 &\propto \frac{16 \Delta \Omega}{\left( \kappa^2+4(\Delta+\Omega)^2 \right)\left( \kappa^2+4(\Delta-\Omega)^2 \right)} \rightarrow 0 \ ,
\end{align}
for $\kappa \gg \Omega$.
Our experiments with $\kappa \gtrsim 2\pi\times\SI{750}{\kilo\hertz} \gg \Omega = 2\pi\times\SI{1}{\kilo\hertz}$ support this statement, as we did not observe systematic differences between red and blue sidebands (cf Fig.~\ref{fig:plusminus}) over the entire parameter space.

\section{Driving the cavity on resonance for high powers}\label{sec:drive_shift}
\begin{figure}
	\centering
	\includegraphics[]{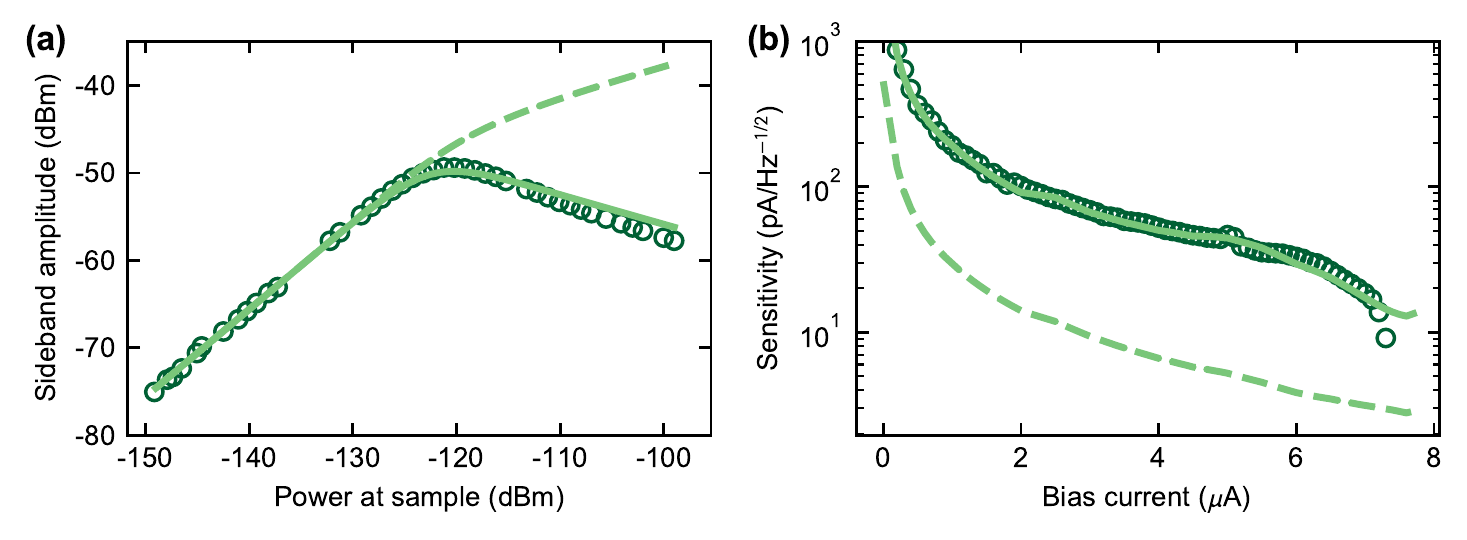}
	\caption{
		\textbf{Driving the cavity on resonance.}
		(a) Data (circles) and model with (solid line) and without (dashed line) detuning of the first order sideband amplitude at $I_0=\SI{4}{\micro\ampere}$, cf. Fig.~\ref{fig:figure3}(c) of the main text.
		Tuning the drive to be matched to the resonance results in a further increased sideband amplitude of more than \SI{10}{\decibel}.
		(b) Data (circles) and model (solid line) of the first order sideband amplitude at $P_\text{in}=\SI{-113}{dBm}$, cf. Fig.~\ref{fig:figure4}(d) of the main text.
		Dashed line corresponds to modelled sensitivity at maximum drive power with shifted drive frequency.
	}
	\label{fig:Duffing-on-res}
\end{figure}

To counteract acquired detuning from the downshift in resonance frequency for high pump powers due to the device nonlinearity, in an ideal measurement configuration the drive tone would also be shifted correspondingly.
As depicted in Fig.~\ref{fig:Duffing-on-res}, in such a situation, the sideband amplitude would keep increasing by more than \SI{10}{\decibel}, resulting in a minimum $\mathcal{S}_I=\SI{2.6}{\pico\ampere\per\sqrt{\hertz}}$.

\section{Deviations between data and theory for high powers}\label{sec:deviation_power}

\begin{figure}
	\centering
	\includegraphics[]{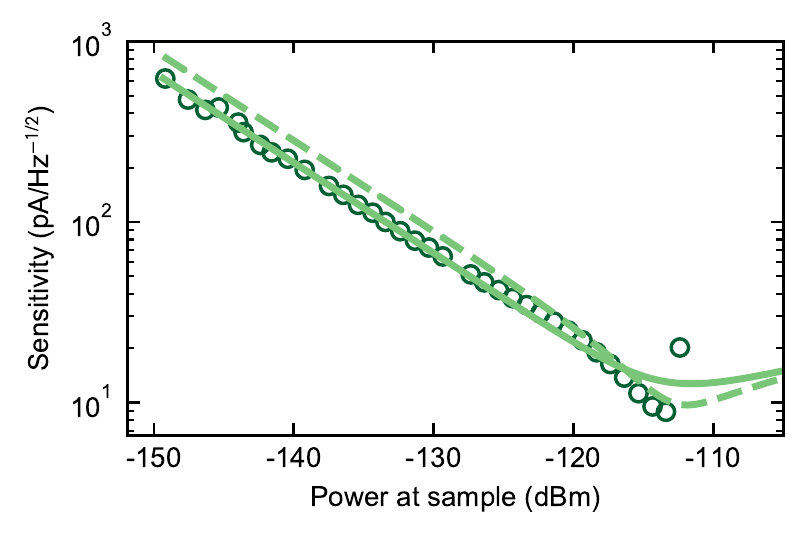}
	\caption{
		\textbf{Deviations between data and theory for high powers.}
		(a) Data (circles) and model (lines) for current sensitivity versus input power for the first order sideband amplitude at $I_0=\SI{7.3}{\micro\ampere}$, cf. Fig.~\ref{fig:figure4}(c) of the main text.
		Solid line corresponds to the same model as in the main text, dashed line has \SI{-600}{\kilo\hertz} extra detuning added, corresponding to an initially red-detuned drive.
		This way, better matching between data and theory is achieved for high pump powers
	}
	\label{fig:deviation_power}
\end{figure}

We observe deviations between the measured and modelled current sensitivity at high drive powers.
As stated in the main text, assuming an initially red-detuned drive, i.e. $\Delta<0$ in the limit of $\abs{\alpha}\rightarrow 0$ could explain this behavior.
In Figure~\ref{fig:deviation_power}, we plot the data with the original model, and add an initial detuning of \SI{-600}{\kilo\hertz} to the drive tune.
While in this case the sensitivity is overestimated for small drive powers, the model follows the measured data closer for high powers than the calculations for zero initial detuning.

\section{Calculating the current sensitivity}\label{sec:analysis}

The current sensitivity is defined by
\begin{align}
\mathcal{S}_I & = \frac{\sigma_I}{\sqrt{\text{ENBW}}}\ ,
\end{align}
with $\sigma_I$ the magnitude of the current noise and ENBW the equivalent noise bandwidth of the spectrum analyzer.
In our experiment, the DUT converts the current modulation $I_\text{LF}$ into an up-converted voltage signal, which we detect as the amplitude of the sidebands as $P_\text{LF}=10^{S/10}$ in \si{\watt}, with $S$ the signal height in \si{dBm}.
Additionally, we record the noise floor amplitude $P_\text{N}=10^{N/10}$ which sets the minimum detectable power, and the signal to noise ratio $\text{SNR}=10^{(S-N)/10}$.
Since the detected power is proportional to the square of the voltage field, which in turn is proportional to the input current, $P \propto V^2 \propto I_\text{LF}^2$, we can infer the equivalent white current noise level of the HEMT and the sensitivity via
\begin{align}
\sigma_I &= \frac{I_\text{LF}}{\sqrt{10^{(S-N)/10}}} \\
\mathcal{S}_I &= \frac{I_\text{LF}}{\sqrt{\text{ENBW}\times10^{(S-N)/10}}}\ .
\end{align}
For a Gaussian filter such as the one used in our setup, $\text{ENBW}=1.065 \times \text{RBW}$, with RBW the resolution bandwidth of the spectrum analyzer which was set to \SI{5}{\hertz} for all measurements \cite{rauscherFundamentalsSpectrumAnalysis2016a}.
In practice, we extract the sideband amplitude from the measured spectra as the peak power value at the expected $\omega_0 \pm m\Omega$ and compute the noise floor as the average of the remaining data points.

\section{Data visualization}

\begin{figure}
	\centering
	\includegraphics[]{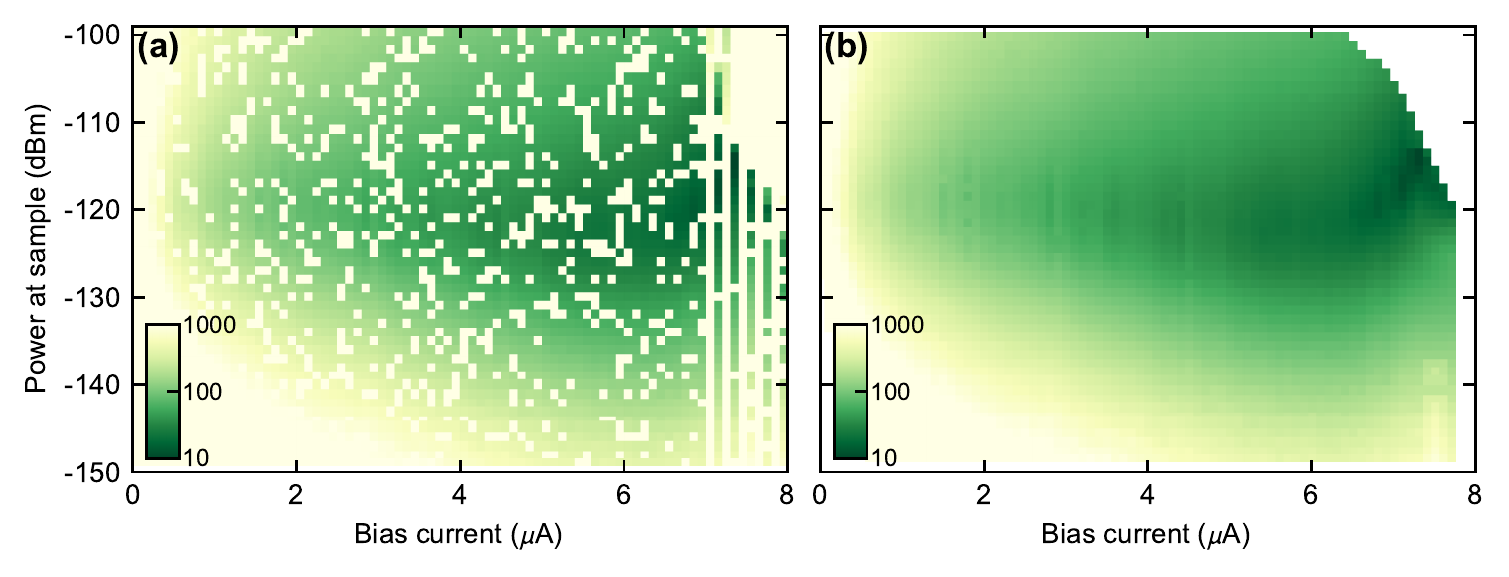}
	\caption{
		\textbf{Interpolating the dataset.}
		Raw (a) and interpolated (b) measured current sensitivities of the $-1\Omega$ sideband for all applied bias currents and pump powers.
		Outliers with large values of $\mathcal{S}_I$ are due to off-resonant drive tones (streaks) or absent LF modulation (speckles) due to errors in the measurement setup, resulting in no mixing at all (see text for details).
		The data in (b) corresponds to Fig.~\ref{fig:figure4}(a) of the main text, with modified colorscale for enhanced visibility.
	}
	\label{fig:fig4aclipnointerp}
\end{figure}

Our raw measurements include a significant number of outliers in current sensitivity, visible as bright spots and streaks in Fig.~\ref{fig:fig4aclipnointerp}(a).
These are due to absent sidebands of all integer multiples of $\Omega$, resulting in apparent negligible SNR and $\mathcal{S}_I > 1000$ for these operating points.
The streaks between \SIrange{7}{8}{\micro\ampere} are due to the pump frequency not correctly adjusted to compensate for the shift due to changing bias current, resulting in very large detuning and undetectable sidebands.
For the remaining outliers, the pump was adjusted correctly, yet still no sidebands appear in the measurement spectra.
We attribute this to the AWG output randomly not being turned on, thus no input modulation was applied and no sidebands produced.
To exclude these outliers from further analysis, we chose to discard data points differing by more than \SI{50}{\percent} from the value expected from theory, and subsequently interpolated the missing experimental data from the surrounding remaining data points.

\section{Modeling the Josephson array CPW}\label{sec:optimized}

\begin{figure}[h]
	\centering
	\includegraphics[]{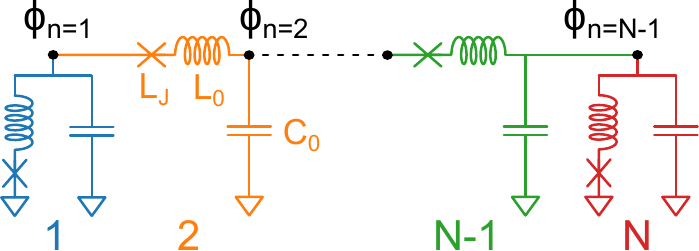}
	\caption{
		\textbf{The JJ array CPW.}
		Circuit schematic of a transmission line resonator consisting of $N$ unit cells based on series and parallel combinations of lumped linear inductors $L_0$, capacitors $C_0$ and Josephson junctions $L_J$, used for deriving an analytical expression for the resonance frequency and anharmonicity.
		The $\phi_n$ indicate the flux at the individual circuit nodes.
		Colors mark different unit cells, with unit cell number indicated below each element group.
		Note that while there are $N$ unit cells, there are only $N-1$ circuit nodes.
	}
	\label{fig:JJCPW-sketch}
\end{figure}

To model a Josephson junction array transmission line resonator, we use $N$ unit cells of length $l$, a transmission line inductance per unit length $L'$, a capacitance per unit length $C'$ and a lumped element Josephson inductance $L_J$, as depicted in Fig.~\ref{fig:JJCPW-sketch} and Fig.~\ref{fig:figure5}(a) of the main text.
Each unit cell has the inductance $L_n = L'l + L_J=L_0+L_J$ and the capacitance $C_n = C'l = C_0$.

\subsection{Full analytical model}\label{sec:analytical}

In order to derive an analytical model, we follow the approach to circuit quantization presented in Ref.~\cite{gelyQuCATQuantumCircuit2019}.
The admittance matrix $\mathbf{Y}$ which relates voltages $v_n$ of a node $n$ to the current injected by a hypothetical infinite impedance source $i_n$ following 
\begin{align}
\mathbf{Y}\mathbf{v} = \mathbf{i}
\label{eq:v_to_i_relation}
\end{align}
explicitly writes
\begin{align}
\underbrace{\begin{pmatrix} 
	2Y_s(\omega)+Y_g (\omega) & - Y_s(\omega) & 0 & \dots & 0 \\
	- Y_s(\omega) & 2Y_s(\omega)+Y_g (\omega) & - Y_s(\omega) & \ddots & \vdots \\
	0 & - Y_s(\omega) & \ddots & \ddots & \vdots  \\
	\vdots & \ddots & \ddots & 2Y_s(\omega)+Y_g (\omega) &- Y_s(\omega) \\
	0 & \dots & \dots & -Y_s(\omega) & 2Y_s(\omega)+Y_g (\omega)
	\end{pmatrix}}_{\mathbf{Y}}
\begin{pmatrix}
v_{1} \\
v_{2} \\
\vdots \\
v_{N-2} \\
v_{N-1}
\end{pmatrix}
=
\begin{pmatrix}
i_{1} \\
i_{2} \\
\vdots \\
i_{N-2} \\
i_{N-1}
\end{pmatrix}
\label{eq:admittance_matrix}
\end{align}

where we have defined the admittances of the series and parallel blocks of the chain by $Y_s(\omega)  = 1/(i \omega (L_J+L_0))$ and $Y_g(\omega)  = i \omega C_0$ respectively.
Such a tridiagonal Toeplitz matrix \cite{noscheseTridiagonalToeplitzMatrices2013} has well known eigenvalues $\zeta_m$ and eigenvectors $e_m$, given here by
\begin{align}
\zeta_m(\omega) &=  2 \left[ 1 - \cos \left(\frac{\pi m}{N}\right) \right] + \frac{Y_g(\omega)}{Y_s (\omega)} \\
e_{m}(n) &= \sqrt{\frac{2}{N}} \sin\left(\frac{\pi m \, n}{N}\right)
\end{align}
for $m\in[1,N-1]$.
Normal mode frequencies $\omega_m$ are those which cancel the determinant of $\textbf{Y}$.
Since the determinant is proportional to the product of eigenvalues $\zeta_m(\omega)$, the frequencies of the $N-1$ modes satisfy $\zeta_m(\omega_m) = 0$,
\begin{align}
\omega_m = \sqrt{\frac{2-2\cos\left(\frac{\pi m}{N}\right)}{(L_J+L_0)C_0}}\ .
\label{eq:anh-wm}
\end{align}

The zero-point fluctuations in flux across the first inductive element (series combination of junction and inductor) for a mode $m$ is determined by the imaginary part of the derivative of the admittance $Y_1 = \left(i_1/v_1\right)_{i_n = 0, n\ne 1}$, evaluated at $\omega_m$.
To obtain $Y_1$, we write the admittance matrix as $\mathbf{Y} = \mathbf{U} \mathbf{D} \mathbf{U}^{T} $ where $  \mathbf{D} $ is the diagonal matrix with  $m$-th diagonal element $\zeta_m$,  and $ \mathbf{U}$ is a matrix whose $m$-th row is $e_m$.
Using this form to invert Eq.~\eqref{eq:v_to_i_relation} leads to
\begin{align}
\left(
\begin{array}{c}
v_{1} \\
v_{2} \\
\dots \\
v_{N-1}
\end{array}
\right)
=
\mathbf{U} \mathbf{D}^{-1} \mathbf{U}^{T}
\left(
\begin{array}{c}
i_1   \\
0     \\
\dots \\
0
\end{array}
\right)
\end{align}
leading to
\begin{align}
\begin{split}
Y_{1}(\omega)&=  \frac{N Y_s (\omega)}{\sum_{m=0}^{N-1}\frac{1}{a_m(\omega)}}\\
a_0(\omega) &= 1\\
a_{m}(\omega) &= \frac{1+2\frac{Y_s(\omega)}{Y_g(\omega)}(1 - \cos (\pi m/ N))}{ 1 + \cos (\pi m/ N) })\text{ for }m>0\ .
\end{split}
\end{align}
To compute its derivative, evaluated at $\omega_m$, we rewrite $Y_1$ as
\begin{align}
Y_{1}(\omega)=  a_m(\omega)\frac{N Y_s(\omega) }{1+\sum_{m'=0}^{N-1}\frac{a_m(\omega)}{a_{m'}(\omega)}\delta_{mm^\prime}}
\end{align}
with $\delta_{mm^\prime}$ the Kronecker delta, such that
\begin{align}
\frac{\partial Y_{1}(\omega)}{\partial\omega}=  \frac{\partial a_m(\omega)}{\partial\omega}\frac{N Y_s(\omega) }{1+\sum_{m'=0}^{N-1}\frac{a_m(\omega)}{a_{m'}(\omega)}\delta_{mm^\prime}}+ a_m(\omega)\frac{\partial }{\partial\omega}\left[\frac{N Y_s(\omega) }{1+\sum_{m'=0}^{N-1}\frac{a_m(\omega)}{a_{m'}(\omega)}\delta_{mm^\prime}}\right]\ .
\end{align}
Since $a_{m'}(\omega_m)\propto \lambda_{m'}(\omega_m) = 0$ if $m'\ne m$, evaluating the derivative at $\omega_m$ and taking its imaginary part yields
\begin{align}
\begin{split}
\text{Im}Y_1'(\omega_m)&=\text{Im}\left(\left.\frac{\partial Y_{1}(\omega)}{\partial\omega}\right|_{\omega = \omega_m}\right)\\
&=  \text{Im}\left(NY_s(\omega_m)\left. \frac{\partial a_m(\omega)}{\partial\omega}\right|_{\omega = \omega_m}\right)\\
&=\frac{N C_0}{1-\cos^2 (\pi m/ N)}\ .
\label{eq:anh-y1p}
\end{split}
\end{align}
The zero-point fluctuations in flux across the first inductive elements for a mode $m$ is then given by \cite{gelyQuCATQuantumCircuit2019,niggBlackBoxSuperconductingCircuit2012}
\begin{align}
\phi_{\text{zpf},1,m} = \sqrt{\frac{\hbar}{\omega_m~\text{Im}Y_1'(\omega_m)}}\ .
\end{align}
The definition of flux \cite{vool_introductionquantum_2017} $\phi_n(t) = \int_{-\infty}^tv_n(\tau)d\tau$ translates in the frequency domain to $\phi_n(\omega) = i\omega v_n(\omega)$.
So knowing the relation between the node voltage amplitudes at a frequency $\omega_m$, given by the coefficients $e_m(n)$, is sufficient to convert the fluctuations in flux at the first node to another.
We are interested in the fluctuations in flux across the $n$th inductive element which is given by
\begin{align}
\phi_{\text{zpf},n,m} = \phi_{\text{zpf},1,m}\frac{e_m(n)-e_m(n-1)}{e_m(1)}
\end{align}
for $n\in[1,N]$.
The fluctuations in flux across the $n$th junction are then
\begin{align}
\left(\frac{L_J}{L_J+L_0}\right)\phi_{\text{zpf},n,m}
\end{align}
This leads to the total anharmonicity $A_m$ for a mode $m$
\begin{align}
A_m = \frac{1}{2\phi_0^2L_J}\left(\frac{L_J}{L_J+L_0}\right)^4\sum_{n=1}^{N}\phi_{\text{zpf},n,m}^4
\label{eq:anh-qucat}
\end{align}
where $\phi_0 = \hbar/2e$ is the reduced flux quantum.

\begin{figure}
	\centering
	\includegraphics[]{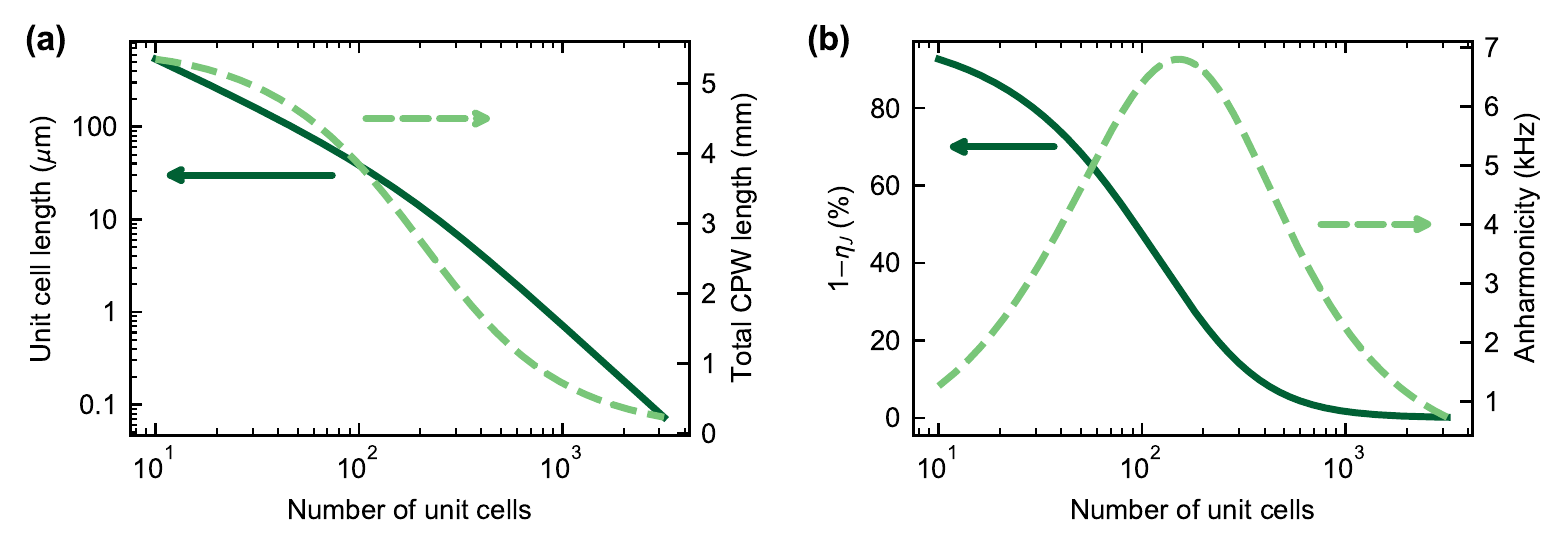}
	\caption{
		\textbf{Device parameters for the JJ CPW.}
		(a) Unit cell length $l$ (solid, left) and total CPW length $l_0$ (dashed, right) for varying unit cell number $N$.
		(b) Normal inductance per unit cell (solid, left) and device anharmonicity (dashed, right) for varying unit cell number $N$.
		All quantities are calculated with the full analytical model for a resonance frequency at \SI{7.5}{\giga\hertz}.
	}
	\label{fig:JJCPW}
\end{figure}

Given an initial resonance frequency for zero bias current of \SI{7.5}{\giga\hertz} and the CPW parameters as specified in the main text for a \SI{1}{\micro\meter} wide CPW, we can use Eq.~\eqref{eq:anh-wm} to calculate the relation between unit cell length and number of unit cells, cf. Fig.\ref{fig:JJCPW}(a).
Compared to the device in the main text, the JJ CPW can be significantly shorter, e.g. $l_0=\SI{845}{\micro\meter}$ for a unit cell length $l=\SI{1}{\micro\meter}$.
The higher the number of unit cells, the shorter the individual unit cells, which leads to an increase in the participation ratio of the Josephson inductance to the total inductance per unit cell, 
\begin{align}
\eta_J = \frac{L_J}{L_J+L_0} \ ,
\end{align}
and the smaller the contribution of normal inductance $1-\eta_J$, cf. Fig.\ref{fig:JJCPW}(b).
The anharmonicity has a maximum of \SI{6.8}{\kilo\hertz} for a unit cell number $N=154$ which corresponds to $\eta_J=2/3$, but drops rapidly for larger $N$.

\begin{figure}
	\centering
	\includegraphics[]{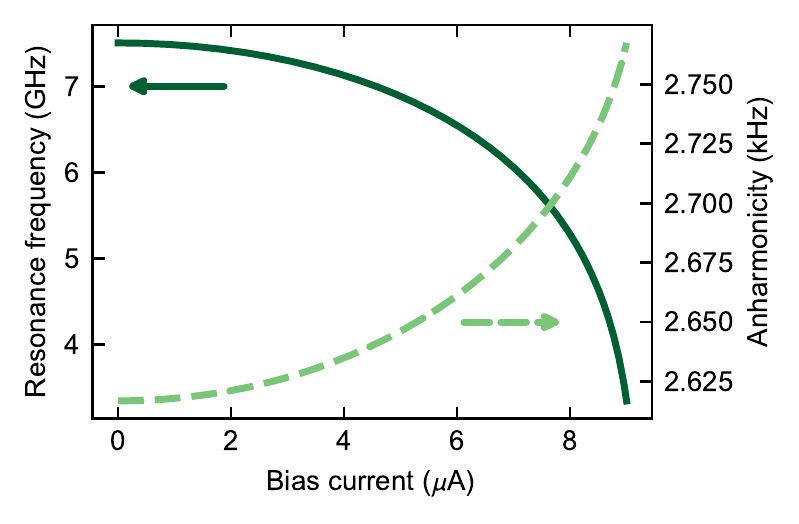}
	\caption{
		\textbf{Bias current tuning of the \SI{1}{\micro\meter} JJ CPW.}
		Solid line: Resonance frequency versus bias current for the proposed JJ CPW device.
		Dashed line: Anharmonicity versus bias current for the proposed JJ CPW device.
		Arrows indicate corresponding axes.
	}
	\label{fig:JJCPW-anh}
\end{figure}

Motivated by a larger current responsivity for large $\eta_J$, our proposed device has 845 unit cells and $\eta_J=\SI{97.7}{\percent}$.
All parameters are detailed in Tab.~\ref{tab:arraygeometry}.
We plot the calculated resonance frequency and anharmonicity for the proposed device design in Fig.~\ref{fig:figure5} from the main text as a function of bias current in Fig.~\ref{fig:JJCPW-anh}.
Since Josephson inductance dominates, resulting in the resonance frequency tuning by more than \SI{55}{\percent}.

\begin{table}
	\caption{Device parameters for the proposed JJ CPW\label{tab:arraygeometry}}
	\begin{tabular}{ccc}
		\hline \hline
		Symbol       & Description                           & Value                            \\
		\hline
		$s$          & CPW center conductor                  & \SI{1}{\micro\meter}            \\
		$w$          & CPW gaps to ground                    & \SI{0.6}{\micro\meter}             \\
		$t$          & Base layer thickness                  & \SI{80}{\nano\meter}             \\
		$l$          & Unit cell length                  & \SI{1}{\micro\meter}             \\
		$N$          & Number of unit cells                 & 845             \\
		$l_0$        & total CPW length   & \SI{845}{\micro\meter}          \\
		$L_J$ & Josephson inductance per unit cell         & \SI{35.9}{\pico\henry} \\
		$L_0$ & Normal inductance per unit cell         & \SI{842}{\femto\henry} \\
		$C_0$ & Geometric capacitance per unit cell      & \SI{169}{\atto\farad}  \\
		\hline\hline
	\end{tabular}
\end{table}

\subsection{Analytical model in the limit of large $N$}\label{sec:analytical-largeN}

We now study the fundamental mode ($m=1$) of an array with many unit-cells ($N\gg 1$).
By Taylor expanding the cosine of Eqs.~\eqref{eq:anh-wm} and \eqref{eq:anh-y1p}, the fundamental mode frequency and derivative of the admittance are then given by
\begin{align}
\omega_1 &\simeq \frac{\pi}{N\sqrt{C_0(L_J+L_0)}} \label{eq:omega-limit}\\
\text{Im}Y_1'(\omega_1) &\simeq\frac{N^3}{\pi^2}C_0\ .
\end{align}
The quantity which relates zero-point fluctuations in phase accross the $n$th unit cell to the zero-point fluctuations of the first unit-cell can be simplified to
\begin{align}
\frac{e_m(n)-e_m(n-1)}{e_m(1)}\simeq \frac{\sin(\frac{\pi n}{N})-\sin(\frac{\pi n}{N}-\frac{\pi}{N})}{\frac{\pi}{N}}\simeq \cos\left(\frac{\pi n}{N}\right)
\end{align}
Plugging these quantities into the expression of the anharmonicity, leads to
\begin{align}
A_1 &\simeq \frac{\hbar^2}{2\phi_0^2L_J }\frac{N^2C_0(L_J+L_0)}{\pi^2}\left(\frac{L_J}{L_J+L_0}\right)^4 \frac{\pi^2}{N^4C_0^2} \sum_{n=1}^{N}\cos^4\left(\frac{\pi n}{N}\right) \\
&= \frac{3\pi^2}{4N^3}\left(\frac{L_J}{L_J+L_0}\right)^3 \frac{e^2}{C_0}
\label{eq:anh-limit}
\end{align}
where $\phi_0 = \hbar/2e$ is the reduced flux quantum and we made use of the relation $\sum_{n=1}^N \cos(n\pi/N)^4=3N/8$.

\subsection{Alternative derivation}

We assume that the fundamental cavity mode $m=1$ of the JJ array CPW has current antinodes at both ends, i.e. we are dealing with a $\lambda/2$ cavity such that the resonator length $l_1 = \lambda_1/2$ with the resonance wavelength $\lambda_1$.
The resonance frequency of the fundamental mode dependent on $N$ and $l$ is given by
\begin{align}
\omega_1 = \frac{\pi}{N\sqrt{C'l\left(L'l + L_J\right)}} \ ,
\end{align}
which is equivalent to Eq.~\eqref{eq:omega-limit}.
For given $C'$, $L'$, $L_J$, $\omega_1$ and $N$, this allows for the calculation of the needed unit cell length $l$.

As we are working with a half-wavelength mode, the basic relation between the resonance frequency and the zero-point fluctuation flux per length in the limit of a continuous flux distribution is given by
\begin{align}
\frac{1}{2}\hbar \omega_1 = \int_0^{\lambda_1/2} \frac{\Phi_\mathrm{zpf}'^2}{L_n'} dx
\end{align}
where $L_n' = L_n/l$ and $\Phi_\mathrm{zpf}' = \Phi_z'\cos\left(\frac{2\pi}{\lambda_1}x\right)$ is the flux per length of transmission line.
This corresponds to
\begin{align}
\Phi_z' = \sqrt{\frac{\hbar \omega_1 L_n'}{l_1}}\ .
\end{align}
Hence, the flux of the $n$th junction is approximately given by
\begin{align}
\Phi_n = \frac{L_J}{L_n}\Phi_z' l \cos{\left(\frac{2\pi}{\lambda_1}\left[n-\frac{1}{2}\right]l\right)}
\end{align}
where the first factor takes into account that only part of the flux is across the junction.
With the Josephson energy $E_J = \frac{\Phi_0^2}{4\pi^2 L_J}$, the anharmonicity is given by
\begin{align}
A &= \frac{12\pi^2}{6L_J\Phi_0^2}\sum_{n=1}^N \Phi_n^4 \\
&= \frac{2e^2 \omega_1^2}{N^2} \frac{L_J^3}{L_n^2}\sum_{n=1}^{N}\cos^4\left(\frac{\pi}{N}\left[n-\frac{1}{2}\right] \right) \\
&= \frac{3\pi^2}{4N^3}\left(\frac{L_J}{L_J+L_0}\right)^3 \frac{e^2}{C_0} \ ,
\label{eq:anh-closed}
\end{align}
where we have used the fact that the cosine sum for values $N>2$ is equal to $3N/8$, which is identical to the result of the full analytical model in the limit of large $N$, cf. Eq.~\eqref{eq:anh-limit}.


\begin{thebibliography}{40}%
\makeatletter
\providecommand \@ifxundefined [1]{%
 \@ifx{#1\undefined}
}%
\providecommand \@ifnum [1]{%
 \ifnum #1\expandafter \@firstoftwo
 \else \expandafter \@secondoftwo
 \fi
}%
\providecommand \@ifx [1]{%
 \ifx #1\expandafter \@firstoftwo
 \else \expandafter \@secondoftwo
 \fi
}%
\providecommand \natexlab [1]{#1}%
\providecommand \enquote  [1]{``#1''}%
\providecommand \bibnamefont  [1]{#1}%
\providecommand \bibfnamefont [1]{#1}%
\providecommand \citenamefont [1]{#1}%
\providecommand \href@noop [0]{\@secondoftwo}%
\providecommand \href [0]{\begingroup \@sanitize@url \@href}%
\providecommand \@href[1]{\@@startlink{#1}\@@href}%
\providecommand \@@href[1]{\endgroup#1\@@endlink}%
\providecommand \@sanitize@url [0]{\catcode `\\12\catcode `\$12\catcode
  `\&12\catcode `\#12\catcode `\^12\catcode `\_12\catcode `\%12\relax}%
\providecommand \@@startlink[1]{}%
\providecommand \@@endlink[0]{}%
\providecommand \url  [0]{\begingroup\@sanitize@url \@url }%
\providecommand \@url [1]{\endgroup\@href {#1}{\urlprefix }}%
\providecommand \urlprefix  [0]{URL }%
\providecommand \Eprint [0]{\href }%
\providecommand \doibase [0]{http://dx.doi.org/}%
\providecommand \selectlanguage [0]{\@gobble}%
\providecommand \bibinfo  [0]{\@secondoftwo}%
\providecommand \bibfield  [0]{\@secondoftwo}%
\providecommand \translation [1]{[#1]}%
\providecommand \BibitemOpen [0]{}%
\providecommand \bibitemStop [0]{}%
\providecommand \bibitemNoStop [0]{.\EOS\space}%
\providecommand \EOS [0]{\spacefactor3000\relax}%
\providecommand \BibitemShut  [1]{\csname bibitem#1\endcsname}%
\let\auto@bib@innerbib\@empty
\bibitem [{\citenamefont {Goldie}\ \emph {et~al.}(2011)\citenamefont {Goldie},
  \citenamefont {Velichko}, \citenamefont {Glowacka},\ and\ \citenamefont
  {Withington}}]{goldieUltralownoiseMoCuTransition2011}%
  \BibitemOpen
  \bibfield  {author} {\bibinfo {author} {\bibfnamefont {D.~J.}\ \bibnamefont
  {Goldie}}, \bibinfo {author} {\bibfnamefont {A.~V.}\ \bibnamefont
  {Velichko}}, \bibinfo {author} {\bibfnamefont {D.~M.}\ \bibnamefont
  {Glowacka}}, \ and\ \bibinfo {author} {\bibfnamefont {S.}~\bibnamefont
  {Withington}},\ }\bibfield  {title} {\enquote {\bibinfo {title}
  {Ultra-low-noise {{MoCu}} transition edge sensors for space applications},}\
  }\href {\doibase 10.1063/1.3561432} {\bibfield  {journal} {\bibinfo
  {journal} {Journal of Applied Physics}\ }\textbf {\bibinfo {volume} {109}},\
  \bibinfo {pages} {084507} (\bibinfo {year} {2011})}\BibitemShut {NoStop}%
\bibitem [{\citenamefont {Cabrera}\ \emph {et~al.}(1998)\citenamefont
  {Cabrera}, \citenamefont {Clarke}, \citenamefont {Colling}, \citenamefont
  {Miller}, \citenamefont {Nam},\ and\ \citenamefont
  {Romani}}]{cabreraDetectionSingleInfrared1998}%
  \BibitemOpen
  \bibfield  {author} {\bibinfo {author} {\bibfnamefont {B.}~\bibnamefont
  {Cabrera}}, \bibinfo {author} {\bibfnamefont {R.~M.}\ \bibnamefont {Clarke}},
  \bibinfo {author} {\bibfnamefont {P.}~\bibnamefont {Colling}}, \bibinfo
  {author} {\bibfnamefont {A.~J.}\ \bibnamefont {Miller}}, \bibinfo {author}
  {\bibfnamefont {S.}~\bibnamefont {Nam}}, \ and\ \bibinfo {author}
  {\bibfnamefont {R.~W.}\ \bibnamefont {Romani}},\ }\bibfield  {title}
  {\enquote {\bibinfo {title} {Detection of single infrared, optical, and
  ultraviolet photons using superconducting transition edge sensors},}\ }\href
  {\doibase 10.1063/1.121984} {\bibfield  {journal} {\bibinfo  {journal}
  {Applied Physics Letters}\ }\textbf {\bibinfo {volume} {73}},\ \bibinfo
  {pages} {735--737} (\bibinfo {year} {1998})}\BibitemShut {NoStop}%
\bibitem [{\citenamefont {Miller}\ \emph {et~al.}(2003)\citenamefont {Miller},
  \citenamefont {Nam}, \citenamefont {Martinis},\ and\ \citenamefont
  {Sergienko}}]{millerDemonstrationLownoiseNearinfrared2003}%
  \BibitemOpen
  \bibfield  {author} {\bibinfo {author} {\bibfnamefont {A.~J.}\ \bibnamefont
  {Miller}}, \bibinfo {author} {\bibfnamefont {S.~W.}\ \bibnamefont {Nam}},
  \bibinfo {author} {\bibfnamefont {J.~M.}\ \bibnamefont {Martinis}}, \ and\
  \bibinfo {author} {\bibfnamefont {A.~V.}\ \bibnamefont {Sergienko}},\
  }\bibfield  {title} {\enquote {\bibinfo {title} {Demonstration of a low-noise
  near-infrared photon counter with multiphoton discrimination},}\ }\href
  {\doibase 10.1063/1.1596723} {\bibfield  {journal} {\bibinfo  {journal}
  {Applied Physics Letters}\ }\textbf {\bibinfo {volume} {83}},\ \bibinfo
  {pages} {791--793} (\bibinfo {year} {2003})}\BibitemShut {NoStop}%
\bibitem [{\citenamefont {Gay}\ \emph {et~al.}(2000)\citenamefont {Gay},
  \citenamefont {Piquemal},\ and\ \citenamefont
  {Genev{\`e}s}}]{gayUltralowNoiseCurrent2000}%
  \BibitemOpen
  \bibfield  {author} {\bibinfo {author} {\bibfnamefont {F.}~\bibnamefont
  {Gay}}, \bibinfo {author} {\bibfnamefont {F.}~\bibnamefont {Piquemal}}, \
  and\ \bibinfo {author} {\bibfnamefont {G.}~\bibnamefont {Genev{\`e}s}},\
  }\bibfield  {title} {\enquote {\bibinfo {title} {Ultralow noise current
  amplifier based on a cryogenic current comparator},}\ }\href {\doibase
  10.1063/1.1326054} {\bibfield  {journal} {\bibinfo  {journal} {Review of
  Scientific Instruments}\ }\textbf {\bibinfo {volume} {71}},\ \bibinfo {pages}
  {4592--4595} (\bibinfo {year} {2000})}\BibitemShut {NoStop}%
\bibitem [{\citenamefont {Henderson}\ \emph {et~al.}(2016)\citenamefont
  {Henderson}, \citenamefont {Stevens}, \citenamefont {Amiri}, \citenamefont
  {Austermann}, \citenamefont {Beall}, \citenamefont {Chaudhuri}, \citenamefont
  {Cho}, \citenamefont {Choi}, \citenamefont {Cothard}, \citenamefont
  {Crowley}, \citenamefont {Duff}, \citenamefont {Fitzgerald}, \citenamefont
  {Gallardo}, \citenamefont {Halpern}, \citenamefont {Hasselfield},
  \citenamefont {Hilton}, \citenamefont {Ho}, \citenamefont {Hubmayr},
  \citenamefont {Irwin}, \citenamefont {Koopman}, \citenamefont {Li},
  \citenamefont {Li}, \citenamefont {McMahon}, \citenamefont {Nati},
  \citenamefont {Niemack}, \citenamefont {Reintsema}, \citenamefont {Salatino},
  \citenamefont {Schillaci}, \citenamefont {Schmitt}, \citenamefont {Simon},
  \citenamefont {Staggs}, \citenamefont {Vavagiakis},\ and\ \citenamefont
  {Ward}}]{hendersonReadoutTwokilopixelTransitionedge2016}%
  \BibitemOpen
  \bibfield  {author} {\bibinfo {author} {\bibfnamefont {S.~W.}\ \bibnamefont
  {Henderson}}, \bibinfo {author} {\bibfnamefont {J.~R.}\ \bibnamefont
  {Stevens}}, \bibinfo {author} {\bibfnamefont {M.}~\bibnamefont {Amiri}},
  \bibinfo {author} {\bibfnamefont {J.}~\bibnamefont {Austermann}}, \bibinfo
  {author} {\bibfnamefont {J.~A.}\ \bibnamefont {Beall}}, \bibinfo {author}
  {\bibfnamefont {S.}~\bibnamefont {Chaudhuri}}, \bibinfo {author}
  {\bibfnamefont {H.-M.}\ \bibnamefont {Cho}}, \bibinfo {author} {\bibfnamefont
  {S.~K.}\ \bibnamefont {Choi}}, \bibinfo {author} {\bibfnamefont {N.~F.}\
  \bibnamefont {Cothard}}, \bibinfo {author} {\bibfnamefont {K.~T.}\
  \bibnamefont {Crowley}}, \bibinfo {author} {\bibfnamefont {S.~M.}\
  \bibnamefont {Duff}}, \bibinfo {author} {\bibfnamefont {C.~P.}\ \bibnamefont
  {Fitzgerald}}, \bibinfo {author} {\bibfnamefont {P.~A.}\ \bibnamefont
  {Gallardo}}, \bibinfo {author} {\bibfnamefont {M.}~\bibnamefont {Halpern}},
  \bibinfo {author} {\bibfnamefont {M.}~\bibnamefont {Hasselfield}}, \bibinfo
  {author} {\bibfnamefont {G.}~\bibnamefont {Hilton}}, \bibinfo {author}
  {\bibfnamefont {S.-P.~P.}\ \bibnamefont {Ho}}, \bibinfo {author}
  {\bibfnamefont {J.}~\bibnamefont {Hubmayr}}, \bibinfo {author} {\bibfnamefont
  {K.~D.}\ \bibnamefont {Irwin}}, \bibinfo {author} {\bibfnamefont {B.~J.}\
  \bibnamefont {Koopman}}, \bibinfo {author} {\bibfnamefont {D.}~\bibnamefont
  {Li}}, \bibinfo {author} {\bibfnamefont {Y.}~\bibnamefont {Li}}, \bibinfo
  {author} {\bibfnamefont {J.}~\bibnamefont {McMahon}}, \bibinfo {author}
  {\bibfnamefont {F.}~\bibnamefont {Nati}}, \bibinfo {author} {\bibfnamefont
  {M.}~\bibnamefont {Niemack}}, \bibinfo {author} {\bibfnamefont {C.~D.}\
  \bibnamefont {Reintsema}}, \bibinfo {author} {\bibfnamefont {M.}~\bibnamefont
  {Salatino}}, \bibinfo {author} {\bibfnamefont {A.}~\bibnamefont {Schillaci}},
  \bibinfo {author} {\bibfnamefont {B.~L.}\ \bibnamefont {Schmitt}}, \bibinfo
  {author} {\bibfnamefont {S.~M.}\ \bibnamefont {Simon}}, \bibinfo {author}
  {\bibfnamefont {S.~T.}\ \bibnamefont {Staggs}}, \bibinfo {author}
  {\bibfnamefont {E.~M.}\ \bibnamefont {Vavagiakis}}, \ and\ \bibinfo {author}
  {\bibfnamefont {J.~T.}\ \bibnamefont {Ward}},\ }\bibfield  {title} {\enquote
  {\bibinfo {title} {Readout of two-kilopixel transition-edge sensor arrays for
  {{Advanced ACTPol}}},}\ }in\ \href {\doibase 10.1117/12.2233895} {\emph
  {\bibinfo {booktitle} {{{SPIE Astronomical Telescopes}} +
  {{Instrumentation}}}}},\ \bibinfo {editor} {edited by\ \bibinfo {editor}
  {\bibfnamefont {W.~S.}\ \bibnamefont {Holland}}\ and\ \bibinfo {editor}
  {\bibfnamefont {J.}~\bibnamefont {Zmuidzinas}}}\ (\bibinfo {address}
  {{Edinburgh, United Kingdom}},\ \bibinfo {year} {2016})\ p.\ \bibinfo {pages}
  {99141G}\BibitemShut {NoStop}%
\bibitem [{\citenamefont {Kher}\ \emph {et~al.}(2016)\citenamefont {Kher},
  \citenamefont {Day}, \citenamefont {Eom}, \citenamefont {Zmuidzinas},\ and\
  \citenamefont {Leduc}}]{kherKineticInductanceParametric2016}%
  \BibitemOpen
  \bibfield  {author} {\bibinfo {author} {\bibfnamefont {A.}~\bibnamefont
  {Kher}}, \bibinfo {author} {\bibfnamefont {P.~K.}\ \bibnamefont {Day}},
  \bibinfo {author} {\bibfnamefont {B.~H.}\ \bibnamefont {Eom}}, \bibinfo
  {author} {\bibfnamefont {J.}~\bibnamefont {Zmuidzinas}}, \ and\ \bibinfo
  {author} {\bibfnamefont {H.~G.}\ \bibnamefont {Leduc}},\ }\bibfield  {title}
  {\enquote {\bibinfo {title} {Kinetic {{Inductance Parametric
  Up}}-{{Converter}}},}\ }\href {\doibase 10.1007/s10909-015-1364-0} {\bibfield
   {journal} {\bibinfo  {journal} {Journal of Low Temperature Physics}\
  }\textbf {\bibinfo {volume} {184}},\ \bibinfo {pages} {480--485} (\bibinfo
  {year} {2016})}\BibitemShut {NoStop}%
\bibitem [{\citenamefont {Doerner}\ \emph {et~al.}(2018)\citenamefont
  {Doerner}, \citenamefont {Kuzmin}, \citenamefont {Graf}, \citenamefont
  {Charaev}, \citenamefont {Wuensch},\ and\ \citenamefont
  {Siegel}}]{doernerCompactMicrowaveKinetic2018}%
  \BibitemOpen
  \bibfield  {author} {\bibinfo {author} {\bibfnamefont {S.}~\bibnamefont
  {Doerner}}, \bibinfo {author} {\bibfnamefont {A.}~\bibnamefont {Kuzmin}},
  \bibinfo {author} {\bibfnamefont {K.}~\bibnamefont {Graf}}, \bibinfo {author}
  {\bibfnamefont {I.}~\bibnamefont {Charaev}}, \bibinfo {author} {\bibfnamefont
  {S.}~\bibnamefont {Wuensch}}, \ and\ \bibinfo {author} {\bibfnamefont
  {M.}~\bibnamefont {Siegel}},\ }\bibfield  {title} {\enquote {\bibinfo {title}
  {Compact microwave kinetic inductance nanowire galvanometer for cryogenic
  detectors at 4.2 {{K}}},}\ }\href {\doibase 10.1088/2399-6528/aaaa8e}
  {\bibfield  {journal} {\bibinfo  {journal} {Journal of Physics
  Communications}\ }\textbf {\bibinfo {volume} {2}},\ \bibinfo {pages} {025016}
  (\bibinfo {year} {2018})}\BibitemShut {NoStop}%
\bibitem [{\citenamefont {Kuzmin}\ \emph {et~al.}(2018)\citenamefont {Kuzmin},
  \citenamefont {Doerner}, \citenamefont {Singer}, \citenamefont {Charaev},
  \citenamefont {Ilin}, \citenamefont {Wuensch},\ and\ \citenamefont
  {Siegel}}]{kuzminTerahertzTransitionEdgeSensor2018}%
  \BibitemOpen
  \bibfield  {author} {\bibinfo {author} {\bibfnamefont {A.}~\bibnamefont
  {Kuzmin}}, \bibinfo {author} {\bibfnamefont {S.}~\bibnamefont {Doerner}},
  \bibinfo {author} {\bibfnamefont {S.}~\bibnamefont {Singer}}, \bibinfo
  {author} {\bibfnamefont {I.}~\bibnamefont {Charaev}}, \bibinfo {author}
  {\bibfnamefont {K.}~\bibnamefont {Ilin}}, \bibinfo {author} {\bibfnamefont
  {S.}~\bibnamefont {Wuensch}}, \ and\ \bibinfo {author} {\bibfnamefont
  {M.}~\bibnamefont {Siegel}},\ }\bibfield  {title} {\enquote {\bibinfo {title}
  {Terahertz {{Transition}}-{{Edge Sensor With Kinetic}}-{{Inductance
  Amplifier}} at 4.2 {{K}}},}\ }\href {\doibase 10.1109/TTHZ.2018.2872413}
  {\bibfield  {journal} {\bibinfo  {journal} {IEEE Transactions on Terahertz
  Science and Technology}\ }\textbf {\bibinfo {volume} {8}},\ \bibinfo {pages}
  {622--629} (\bibinfo {year} {2018})}\BibitemShut {NoStop}%
\bibitem [{\citenamefont {Stehlik}\ \emph {et~al.}(2015)\citenamefont
  {Stehlik}, \citenamefont {Liu}, \citenamefont {Quintana}, \citenamefont
  {Eichler}, \citenamefont {Hartke},\ and\ \citenamefont
  {Petta}}]{stehlikFastChargeSensing2015}%
  \BibitemOpen
  \bibfield  {author} {\bibinfo {author} {\bibfnamefont {J.}~\bibnamefont
  {Stehlik}}, \bibinfo {author} {\bibfnamefont {Y.-Y.}\ \bibnamefont {Liu}},
  \bibinfo {author} {\bibfnamefont {C.~M.}\ \bibnamefont {Quintana}}, \bibinfo
  {author} {\bibfnamefont {C.}~\bibnamefont {Eichler}}, \bibinfo {author}
  {\bibfnamefont {T.~R.}\ \bibnamefont {Hartke}}, \ and\ \bibinfo {author}
  {\bibfnamefont {J.~R.}\ \bibnamefont {Petta}},\ }\bibfield  {title} {\enquote
  {\bibinfo {title} {Fast {{Charge Sensing}} of a {{Cavity}}-{{Coupled Double
  Quantum Dot Using}} a {{Josephson Parametric Amplifier}}},}\ }\href {\doibase
  10.1103/PhysRevApplied.4.014018} {\bibfield  {journal} {\bibinfo  {journal}
  {Physical Review Applied}\ }\textbf {\bibinfo {volume} {4}},\ \bibinfo
  {pages} {014018} (\bibinfo {year} {2015})}\BibitemShut {NoStop}%
\bibitem [{\citenamefont {Pavolotsky}\ \emph {et~al.}(2011)\citenamefont
  {Pavolotsky}, \citenamefont {Dochev},\ and\ \citenamefont
  {Belitsky}}]{pavolotskyAgingAnnealinginducedVariations2011}%
  \BibitemOpen
  \bibfield  {author} {\bibinfo {author} {\bibfnamefont {A.~B.}\ \bibnamefont
  {Pavolotsky}}, \bibinfo {author} {\bibfnamefont {D.}~\bibnamefont {Dochev}},
  \ and\ \bibinfo {author} {\bibfnamefont {V.}~\bibnamefont {Belitsky}},\
  }\bibfield  {title} {\enquote {\bibinfo {title} {Aging- and annealing-induced
  variations in {{Nb}}/{{Al}}\textendash{{AlOx}}/{{Nb}} tunnel junction
  properties},}\ }\href {\doibase 10.1063/1.3532040} {\bibfield  {journal}
  {\bibinfo  {journal} {Journal of Applied Physics}\ }\textbf {\bibinfo
  {volume} {109}},\ \bibinfo {pages} {024502} (\bibinfo {year}
  {2011})}\BibitemShut {NoStop}%
\bibitem [{\citenamefont {Goteti}\ \emph {et~al.}(2019)\citenamefont {Goteti},
  \citenamefont {Denton}, \citenamefont {Krause}, \citenamefont {Stephen},
  \citenamefont {Sellers}, \citenamefont {Sullivan}, \citenamefont {Hamilton},
  \citenamefont {Wynn},\ and\ \citenamefont
  {Tolpygo}}]{gotetiReliabilityStudiesNb2019}%
  \BibitemOpen
  \bibfield  {author} {\bibinfo {author} {\bibfnamefont {U.~S.}\ \bibnamefont
  {Goteti}}, \bibinfo {author} {\bibfnamefont {M.}~\bibnamefont {Denton}},
  \bibinfo {author} {\bibfnamefont {K.}~\bibnamefont {Krause}}, \bibinfo
  {author} {\bibfnamefont {A.}~\bibnamefont {Stephen}}, \bibinfo {author}
  {\bibfnamefont {J.~A.}\ \bibnamefont {Sellers}}, \bibinfo {author}
  {\bibfnamefont {S.}~\bibnamefont {Sullivan}}, \bibinfo {author}
  {\bibfnamefont {M.~C.}\ \bibnamefont {Hamilton}}, \bibinfo {author}
  {\bibfnamefont {A.}~\bibnamefont {Wynn}}, \ and\ \bibinfo {author}
  {\bibfnamefont {S.~K.}\ \bibnamefont {Tolpygo}},\ }\bibfield  {title}
  {\enquote {\bibinfo {title} {Reliability {{Studies}} of
  {{Nb}}/{{AlOx}}/{{Al}}/{{Nb Josephson Junctions Through Accelerated}}-{{Life
  Electrical Stress Testing}}},}\ }\href {\doibase 10.1109/TASC.2019.2904593}
  {\bibfield  {journal} {\bibinfo  {journal} {IEEE Transactions on Applied
  Superconductivity}\ }\textbf {\bibinfo {volume} {29}},\ \bibinfo {pages}
  {1--7} (\bibinfo {year} {2019})}\BibitemShut {NoStop}%
\bibitem [{\citenamefont {Gunnarsson}\ \emph {et~al.}(2013)\citenamefont
  {Gunnarsson}, \citenamefont {Pirkkalainen}, \citenamefont {Li}, \citenamefont
  {Paraoanu}, \citenamefont {Hakonen}, \citenamefont {Sillanp{\"a}{\"a}},\ and\
  \citenamefont {Prunnila}}]{gunnarssonDielectricLossesMultilayer2013}%
  \BibitemOpen
  \bibfield  {author} {\bibinfo {author} {\bibfnamefont {D.}~\bibnamefont
  {Gunnarsson}}, \bibinfo {author} {\bibfnamefont {J.-M.}\ \bibnamefont
  {Pirkkalainen}}, \bibinfo {author} {\bibfnamefont {J.}~\bibnamefont {Li}},
  \bibinfo {author} {\bibfnamefont {G.~S.}\ \bibnamefont {Paraoanu}}, \bibinfo
  {author} {\bibfnamefont {P.}~\bibnamefont {Hakonen}}, \bibinfo {author}
  {\bibfnamefont {M.}~\bibnamefont {Sillanp{\"a}{\"a}}}, \ and\ \bibinfo
  {author} {\bibfnamefont {M.}~\bibnamefont {Prunnila}},\ }\bibfield  {title}
  {\enquote {\bibinfo {title} {Dielectric losses in multi-layer {{Josephson}}
  junction qubits},}\ }\href {\doibase 10/ggdn7p} {\bibfield  {journal}
  {\bibinfo  {journal} {Superconductor Science and Technology}\ }\textbf
  {\bibinfo {volume} {26}},\ \bibinfo {pages} {085010} (\bibinfo {year}
  {2013})}\BibitemShut {NoStop}%
\bibitem [{\citenamefont {Yanai}\ and\ \citenamefont
  {Steele}(2019)}]{yanaiObservationEnhancedCoherence2019}%
  \BibitemOpen
  \bibfield  {author} {\bibinfo {author} {\bibfnamefont {S.}~\bibnamefont
  {Yanai}}\ and\ \bibinfo {author} {\bibfnamefont {G.~A.}\ \bibnamefont
  {Steele}},\ }\bibfield  {title} {\enquote {\bibinfo {title} {Observation of
  enhanced coherence in {{Josephson SQUID}} cavities using a hybrid fabrication
  approach},}\ }\href@noop {} {\bibfield  {journal} {\bibinfo  {journal}
  {arXiv:1911.07119 [cond-mat]}\ } (\bibinfo {year} {2019})},\ \Eprint
  {http://arxiv.org/abs/1911.07119} {arXiv:1911.07119 [cond-mat]} \BibitemShut
  {NoStop}%
\bibitem [{\citenamefont {Wallraff}\ \emph {et~al.}(2004)\citenamefont
  {Wallraff}, \citenamefont {Schuster}, \citenamefont {Blais}, \citenamefont
  {Frunzio}, \citenamefont {Huang}, \citenamefont {Majer}, \citenamefont
  {Kumar}, \citenamefont {Girvin},\ and\ \citenamefont
  {Schoelkopf}}]{wallraffStrongCouplingSingle2004}%
  \BibitemOpen
  \bibfield  {author} {\bibinfo {author} {\bibfnamefont {A.}~\bibnamefont
  {Wallraff}}, \bibinfo {author} {\bibfnamefont {D.~I.}\ \bibnamefont
  {Schuster}}, \bibinfo {author} {\bibfnamefont {A.}~\bibnamefont {Blais}},
  \bibinfo {author} {\bibfnamefont {L.}~\bibnamefont {Frunzio}}, \bibinfo
  {author} {\bibfnamefont {R.-S.}\ \bibnamefont {Huang}}, \bibinfo {author}
  {\bibfnamefont {J.}~\bibnamefont {Majer}}, \bibinfo {author} {\bibfnamefont
  {S.}~\bibnamefont {Kumar}}, \bibinfo {author} {\bibfnamefont {S.~M.}\
  \bibnamefont {Girvin}}, \ and\ \bibinfo {author} {\bibfnamefont {R.~J.}\
  \bibnamefont {Schoelkopf}},\ }\bibfield  {title} {\enquote {\bibinfo {title}
  {Strong coupling of a single photon to a superconducting qubit using circuit
  quantum electrodynamics},}\ }\href {\doibase 10/bq2bn8} {\bibfield  {journal}
  {\bibinfo  {journal} {Nature}\ }\textbf {\bibinfo {volume} {431}},\ \bibinfo
  {pages} {162--167} (\bibinfo {year} {2004})}\BibitemShut {NoStop}%
\bibitem [{\citenamefont {Lecocq}\ \emph {et~al.}(2011)\citenamefont {Lecocq},
  \citenamefont {Pop}, \citenamefont {Peng}, \citenamefont {Matei},
  \citenamefont {Crozes}, \citenamefont {Fournier}, \citenamefont {Naud},
  \citenamefont {Guichard},\ and\ \citenamefont
  {Buisson}}]{lecocqJunctionFabricationShadow2011}%
  \BibitemOpen
  \bibfield  {author} {\bibinfo {author} {\bibfnamefont {F.}~\bibnamefont
  {Lecocq}}, \bibinfo {author} {\bibfnamefont {I.~M.}\ \bibnamefont {Pop}},
  \bibinfo {author} {\bibfnamefont {Z.}~\bibnamefont {Peng}}, \bibinfo {author}
  {\bibfnamefont {I.}~\bibnamefont {Matei}}, \bibinfo {author} {\bibfnamefont
  {T.}~\bibnamefont {Crozes}}, \bibinfo {author} {\bibfnamefont
  {T.}~\bibnamefont {Fournier}}, \bibinfo {author} {\bibfnamefont
  {C.}~\bibnamefont {Naud}}, \bibinfo {author} {\bibfnamefont {W.}~\bibnamefont
  {Guichard}}, \ and\ \bibinfo {author} {\bibfnamefont {O.}~\bibnamefont
  {Buisson}},\ }\bibfield  {title} {\enquote {\bibinfo {title} {Junction
  fabrication by shadow evaporation without a suspended bridge},}\ }\href
  {\doibase 10/cg6ppq} {\bibfield  {journal} {\bibinfo  {journal}
  {Nanotechnology}\ }\textbf {\bibinfo {volume} {22}},\ \bibinfo {pages}
  {315302} (\bibinfo {year} {2011})}\BibitemShut {NoStop}%
\bibitem [{\citenamefont {Bosman}\ \emph {et~al.}(2015)\citenamefont {Bosman},
  \citenamefont {Singh}, \citenamefont {Bruno},\ and\ \citenamefont
  {Steele}}]{bosmanBroadbandArchitectureGalvanically2015a}%
  \BibitemOpen
  \bibfield  {author} {\bibinfo {author} {\bibfnamefont {S.~J.}\ \bibnamefont
  {Bosman}}, \bibinfo {author} {\bibfnamefont {V.}~\bibnamefont {Singh}},
  \bibinfo {author} {\bibfnamefont {A.}~\bibnamefont {Bruno}}, \ and\ \bibinfo
  {author} {\bibfnamefont {G.~A.}\ \bibnamefont {Steele}},\ }\bibfield  {title}
  {\enquote {\bibinfo {title} {Broadband architecture for galvanically
  accessible superconducting microwave resonators},}\ }\href {\doibase
  10/ggdntc} {\bibfield  {journal} {\bibinfo  {journal} {Applied Physics
  Letters}\ }\textbf {\bibinfo {volume} {107}},\ \bibinfo {pages} {192602}
  (\bibinfo {year} {2015})}\BibitemShut {NoStop}%
\bibitem [{\citenamefont {Schmidt}\ \emph {et~al.}(2018)\citenamefont
  {Schmidt}, \citenamefont {Jenkins}, \citenamefont {Watanabe}, \citenamefont
  {Taniguchi},\ and\ \citenamefont
  {Steele}}]{schmidtBallisticGrapheneSuperconducting2018}%
  \BibitemOpen
  \bibfield  {author} {\bibinfo {author} {\bibfnamefont {F.~E.}\ \bibnamefont
  {Schmidt}}, \bibinfo {author} {\bibfnamefont {M.~D.}\ \bibnamefont
  {Jenkins}}, \bibinfo {author} {\bibfnamefont {K.}~\bibnamefont {Watanabe}},
  \bibinfo {author} {\bibfnamefont {T.}~\bibnamefont {Taniguchi}}, \ and\
  \bibinfo {author} {\bibfnamefont {G.~A.}\ \bibnamefont {Steele}},\ }\bibfield
   {title} {\enquote {\bibinfo {title} {A ballistic graphene superconducting
  microwave circuit},}\ }\href {\doibase 10.1038/s41467-018-06595-2} {\bibfield
   {journal} {\bibinfo  {journal} {Nature Communications}\ }\textbf {\bibinfo
  {volume} {9}},\ \bibinfo {pages} {4069} (\bibinfo {year} {2018})}\BibitemShut
  {NoStop}%
\bibitem [{\citenamefont {Vijay}\ \emph {et~al.}(2009)\citenamefont {Vijay},
  \citenamefont {Sau}, \citenamefont {Cohen},\ and\ \citenamefont
  {Siddiqi}}]{vijayOptimizingAnharmonicityNanoscale2009a}%
  \BibitemOpen
  \bibfield  {author} {\bibinfo {author} {\bibfnamefont {R.}~\bibnamefont
  {Vijay}}, \bibinfo {author} {\bibfnamefont {J.~D.}\ \bibnamefont {Sau}},
  \bibinfo {author} {\bibfnamefont {M.~L.}\ \bibnamefont {Cohen}}, \ and\
  \bibinfo {author} {\bibfnamefont {I.}~\bibnamefont {Siddiqi}},\ }\bibfield
  {title} {\enquote {\bibinfo {title} {Optimizing {{Anharmonicity}} in
  {{Nanoscale Weak Link Josephson Junction Oscillators}}},}\ }\href {\doibase
  10.1103/PhysRevLett.103.087003} {\bibfield  {journal} {\bibinfo  {journal}
  {Physical Review Letters}\ }\textbf {\bibinfo {volume} {103}},\ \bibinfo
  {pages} {087003} (\bibinfo {year} {2009})}\BibitemShut {NoStop}%
\bibitem [{\citenamefont {Kennedy}\ \emph {et~al.}(2019)\citenamefont
  {Kennedy}, \citenamefont {Burnett}, \citenamefont {Fenton}, \citenamefont
  {Constantino}, \citenamefont {Warburton}, \citenamefont {Morton},\ and\
  \citenamefont {{Dupont-Ferrier}}}]{kennedyTunableNbSuperconducting2019a}%
  \BibitemOpen
  \bibfield  {author} {\bibinfo {author} {\bibfnamefont {O.}~\bibnamefont
  {Kennedy}}, \bibinfo {author} {\bibfnamefont {J.}~\bibnamefont {Burnett}},
  \bibinfo {author} {\bibfnamefont {J.}~\bibnamefont {Fenton}}, \bibinfo
  {author} {\bibfnamefont {N.}~\bibnamefont {Constantino}}, \bibinfo {author}
  {\bibfnamefont {P.}~\bibnamefont {Warburton}}, \bibinfo {author}
  {\bibfnamefont {J.}~\bibnamefont {Morton}}, \ and\ \bibinfo {author}
  {\bibfnamefont {E.}~\bibnamefont {{Dupont-Ferrier}}},\ }\bibfield  {title}
  {\enquote {\bibinfo {title} {Tunable {{Nb Superconducting Resonator Based}}
  on a {{Constriction Nano}}-{{SQUID Fabricated}} with a {{Ne Focused Ion
  Beam}}},}\ }\href {\doibase 10/ggd5sc} {\bibfield  {journal} {\bibinfo
  {journal} {Physical Review Applied}\ }\textbf {\bibinfo {volume} {11}},\
  \bibinfo {pages} {014006} (\bibinfo {year} {2019})}\BibitemShut {NoStop}%
\bibitem [{\citenamefont {Rodrigues}\ \emph {et~al.}(2019)\citenamefont
  {Rodrigues}, \citenamefont {Bothner},\ and\ \citenamefont
  {Steele}}]{rodriguesCouplingMicrowavePhotons2019a}%
  \BibitemOpen
  \bibfield  {author} {\bibinfo {author} {\bibfnamefont {I.~C.}\ \bibnamefont
  {Rodrigues}}, \bibinfo {author} {\bibfnamefont {D.}~\bibnamefont {Bothner}},
  \ and\ \bibinfo {author} {\bibfnamefont {G.~A.}\ \bibnamefont {Steele}},\
  }\bibfield  {title} {\enquote {\bibinfo {title} {Coupling microwave photons
  to a mechanical resonator using quantum interference},}\ }\href {\doibase
  10.1038/s41467-019-12964-2} {\bibfield  {journal} {\bibinfo  {journal}
  {Nature Communications}\ }\textbf {\bibinfo {volume} {10}},\ \bibinfo {pages}
  {5359} (\bibinfo {year} {2019})}\BibitemShut {NoStop}%
\bibitem [{\citenamefont {Bothner}\ \emph {et~al.}(2019)\citenamefont
  {Bothner}, \citenamefont {Rodrigues},\ and\ \citenamefont
  {Steele}}]{bothnerPhotonPressureStrongCouplingTwo2019}%
  \BibitemOpen
  \bibfield  {author} {\bibinfo {author} {\bibfnamefont {D.}~\bibnamefont
  {Bothner}}, \bibinfo {author} {\bibfnamefont {I.~C.}\ \bibnamefont
  {Rodrigues}}, \ and\ \bibinfo {author} {\bibfnamefont {G.~A.}\ \bibnamefont
  {Steele}},\ }\bibfield  {title} {\enquote {\bibinfo {title}
  {Photon-{{Pressure Strong}}-{{Coupling}} between two {{Superconducting
  Circuits}}},}\ }\href@noop {} {\bibfield  {journal} {\bibinfo  {journal}
  {arXiv:1911.01262 [cond-mat, physics:quant-ph]}\ } (\bibinfo {year}
  {2019})},\ \Eprint {http://arxiv.org/abs/1911.01262} {arXiv:1911.01262
  [cond-mat, physics:quant-ph]} \BibitemShut {NoStop}%
\bibitem [{See()}]{SeeSupplementalMaterial}%
  \BibitemOpen
  \href@noop {} {\enquote {\bibinfo {title} {See {{Supplemental Material}} at
  [{{URL}} will be inserted by publisher] for details on device fabrication,
  device parameters, hysteresis of switching currents, measurement setup, data
  processing and mathematical derivations.}}\ }\BibitemShut {NoStop}%
\bibitem [{\citenamefont
  {Tinkham}(1996)}]{tinkhamIntroductionSuperconductivity1996}%
  \BibitemOpen
  \bibfield  {author} {\bibinfo {author} {\bibfnamefont {M.}~\bibnamefont
  {Tinkham}},\ }\href@noop {} {\emph {\bibinfo {title} {Introduction to
  {{Superconductivity}}}}},\ \bibinfo {edition} {2nd}\ ed.\ (\bibinfo
  {publisher} {{McGraw-Hill, Inc.}},\ \bibinfo {address} {{New York}},\
  \bibinfo {year} {1996})\BibitemShut {NoStop}%
\bibitem [{\citenamefont {Skocpol}\ \emph {et~al.}(1974)\citenamefont
  {Skocpol}, \citenamefont {Beasley},\ and\ \citenamefont
  {Tinkham}}]{skocpolSelfHeatingHotspots1974}%
  \BibitemOpen
  \bibfield  {author} {\bibinfo {author} {\bibfnamefont {W.~J.}\ \bibnamefont
  {Skocpol}}, \bibinfo {author} {\bibfnamefont {M.~R.}\ \bibnamefont
  {Beasley}}, \ and\ \bibinfo {author} {\bibfnamefont {M.}~\bibnamefont
  {Tinkham}},\ }\bibfield  {title} {\enquote {\bibinfo {title} {Self-heating
  hotspots in superconducting thin-film microbridges},}\ }\href {\doibase
  10.1063/1.1663912} {\bibfield  {journal} {\bibinfo  {journal} {Journal of
  Applied Physics}\ }\textbf {\bibinfo {volume} {45}},\ \bibinfo {pages}
  {4054--4066} (\bibinfo {year} {1974})}\BibitemShut {NoStop}%
\bibitem [{\citenamefont {Hazra}\ \emph {et~al.}(2010)\citenamefont {Hazra},
  \citenamefont {Pascal}, \citenamefont {Courtois},\ and\ \citenamefont
  {Gupta}}]{hazraHysteresisSuperconductingShort2010}%
  \BibitemOpen
  \bibfield  {author} {\bibinfo {author} {\bibfnamefont {D.}~\bibnamefont
  {Hazra}}, \bibinfo {author} {\bibfnamefont {L.~M.~A.}\ \bibnamefont
  {Pascal}}, \bibinfo {author} {\bibfnamefont {H.}~\bibnamefont {Courtois}}, \
  and\ \bibinfo {author} {\bibfnamefont {A.~K.}\ \bibnamefont {Gupta}},\
  }\bibfield  {title} {\enquote {\bibinfo {title} {Hysteresis in
  superconducting short weak links and {$\mu$}-{{SQUIDs}}},}\ }\href {\doibase
  10.1103/PhysRevB.82.184530} {\bibfield  {journal} {\bibinfo  {journal}
  {Physical Review B}\ }\textbf {\bibinfo {volume} {82}},\ \bibinfo {pages}
  {184530} (\bibinfo {year} {2010})}\BibitemShut {NoStop}%
\bibitem [{\citenamefont {Kumar}\ \emph {et~al.}(2015)\citenamefont {Kumar},
  \citenamefont {Fournier}, \citenamefont {Courtois}, \citenamefont
  {Winkelmann},\ and\ \citenamefont
  {Gupta}}]{kumarReversibilitySuperconductingNb2015}%
  \BibitemOpen
  \bibfield  {author} {\bibinfo {author} {\bibfnamefont {N.}~\bibnamefont
  {Kumar}}, \bibinfo {author} {\bibfnamefont {T.}~\bibnamefont {Fournier}},
  \bibinfo {author} {\bibfnamefont {H.}~\bibnamefont {Courtois}}, \bibinfo
  {author} {\bibfnamefont {C.~B.}\ \bibnamefont {Winkelmann}}, \ and\ \bibinfo
  {author} {\bibfnamefont {A.~K.}\ \bibnamefont {Gupta}},\ }\bibfield  {title}
  {\enquote {\bibinfo {title} {Reversibility {{Of Superconducting Nb Weak Links
  Driven By The Proximity Effect In A Quantum Interference Device}}},}\ }\href
  {\doibase 10.1103/PhysRevLett.114.157003} {\bibfield  {journal} {\bibinfo
  {journal} {Physical Review Letters}\ }\textbf {\bibinfo {volume} {114}},\
  \bibinfo {pages} {157003} (\bibinfo {year} {2015})}\BibitemShut {NoStop}%
\bibitem [{\citenamefont {Pogorzalek}\ \emph {et~al.}(2017)\citenamefont
  {Pogorzalek}, \citenamefont {Fedorov}, \citenamefont {Zhong}, \citenamefont
  {Goetz}, \citenamefont {Wulschner}, \citenamefont {Fischer}, \citenamefont
  {Eder}, \citenamefont {Xie}, \citenamefont {Inomata}, \citenamefont
  {Yamamoto}, \citenamefont {Nakamura}, \citenamefont {Marx}, \citenamefont
  {Deppe},\ and\ \citenamefont {Gross}}]{pogorzalekHystereticFluxResponse2017}%
  \BibitemOpen
  \bibfield  {author} {\bibinfo {author} {\bibfnamefont {S.}~\bibnamefont
  {Pogorzalek}}, \bibinfo {author} {\bibfnamefont {K.~G.}\ \bibnamefont
  {Fedorov}}, \bibinfo {author} {\bibfnamefont {L.}~\bibnamefont {Zhong}},
  \bibinfo {author} {\bibfnamefont {J.}~\bibnamefont {Goetz}}, \bibinfo
  {author} {\bibfnamefont {F.}~\bibnamefont {Wulschner}}, \bibinfo {author}
  {\bibfnamefont {M.}~\bibnamefont {Fischer}}, \bibinfo {author} {\bibfnamefont
  {P.}~\bibnamefont {Eder}}, \bibinfo {author} {\bibfnamefont {E.}~\bibnamefont
  {Xie}}, \bibinfo {author} {\bibfnamefont {K.}~\bibnamefont {Inomata}},
  \bibinfo {author} {\bibfnamefont {T.}~\bibnamefont {Yamamoto}}, \bibinfo
  {author} {\bibfnamefont {Y.}~\bibnamefont {Nakamura}}, \bibinfo {author}
  {\bibfnamefont {A.}~\bibnamefont {Marx}}, \bibinfo {author} {\bibfnamefont
  {F.}~\bibnamefont {Deppe}}, \ and\ \bibinfo {author} {\bibfnamefont
  {R.}~\bibnamefont {Gross}},\ }\bibfield  {title} {\enquote {\bibinfo {title}
  {Hysteretic {{Flux Response}} and {{Nondegenerate Gain}} of {{Flux}}-{{Driven
  Josephson Parametric Amplifiers}}},}\ }\href {\doibase
  10.1103/PhysRevApplied.8.024012} {\bibfield  {journal} {\bibinfo  {journal}
  {Physical Review Applied}\ }\textbf {\bibinfo {volume} {8}},\ \bibinfo
  {pages} {024012} (\bibinfo {year} {2017})}\BibitemShut {NoStop}%
\bibitem [{\citenamefont {Kautz}\ and\ \citenamefont
  {Martinis}(1990)}]{kautzNoiseaffectedIVCurves1990b}%
  \BibitemOpen
  \bibfield  {author} {\bibinfo {author} {\bibfnamefont {R.~L.}\ \bibnamefont
  {Kautz}}\ and\ \bibinfo {author} {\bibfnamefont {J.~M.}\ \bibnamefont
  {Martinis}},\ }\bibfield  {title} {\enquote {\bibinfo {title} {Noise-affected
  {{I}}-{{V}} curves in small hysteretic {{Josephson}} junctions},}\ }\href
  {\doibase 10.1103/PhysRevB.42.9903} {\bibfield  {journal} {\bibinfo
  {journal} {Physical Review B}\ }\textbf {\bibinfo {volume} {42}},\ \bibinfo
  {pages} {9903--9937} (\bibinfo {year} {1990})}\BibitemShut {NoStop}%
\bibitem [{\citenamefont {Annunziata}\ \emph {et~al.}(2010)\citenamefont
  {Annunziata}, \citenamefont {Santavicca}, \citenamefont {Frunzio},
  \citenamefont {Catelani}, \citenamefont {Rooks}, \citenamefont {Frydman},\
  and\ \citenamefont
  {Prober}}]{annunziataTunableSuperconductingNanoinductors2010}%
  \BibitemOpen
  \bibfield  {author} {\bibinfo {author} {\bibfnamefont {A.~J.}\ \bibnamefont
  {Annunziata}}, \bibinfo {author} {\bibfnamefont {D.~F.}\ \bibnamefont
  {Santavicca}}, \bibinfo {author} {\bibfnamefont {L.}~\bibnamefont {Frunzio}},
  \bibinfo {author} {\bibfnamefont {G.}~\bibnamefont {Catelani}}, \bibinfo
  {author} {\bibfnamefont {M.~J.}\ \bibnamefont {Rooks}}, \bibinfo {author}
  {\bibfnamefont {A.}~\bibnamefont {Frydman}}, \ and\ \bibinfo {author}
  {\bibfnamefont {D.~E.}\ \bibnamefont {Prober}},\ }\bibfield  {title}
  {\enquote {\bibinfo {title} {Tunable superconducting nanoinductors},}\ }\href
  {\doibase 10/d6t97q} {\bibfield  {journal} {\bibinfo  {journal}
  {Nanotechnology}\ }\textbf {\bibinfo {volume} {21}},\ \bibinfo {pages}
  {445202} (\bibinfo {year} {2010})}\BibitemShut {NoStop}%
\bibitem [{\citenamefont {Vissers}\ \emph {et~al.}(2015)\citenamefont
  {Vissers}, \citenamefont {Hubmayr}, \citenamefont {Sandberg}, \citenamefont
  {Chaudhuri}, \citenamefont {Bockstiegel},\ and\ \citenamefont
  {Gao}}]{vissersFrequencytunableSuperconductingResonators2015}%
  \BibitemOpen
  \bibfield  {author} {\bibinfo {author} {\bibfnamefont {M.~R.}\ \bibnamefont
  {Vissers}}, \bibinfo {author} {\bibfnamefont {J.}~\bibnamefont {Hubmayr}},
  \bibinfo {author} {\bibfnamefont {M.}~\bibnamefont {Sandberg}}, \bibinfo
  {author} {\bibfnamefont {S.}~\bibnamefont {Chaudhuri}}, \bibinfo {author}
  {\bibfnamefont {C.}~\bibnamefont {Bockstiegel}}, \ and\ \bibinfo {author}
  {\bibfnamefont {J.}~\bibnamefont {Gao}},\ }\bibfield  {title} {\enquote
  {\bibinfo {title} {Frequency-tunable superconducting resonators via nonlinear
  kinetic inductance},}\ }\href {\doibase 10.1063/1.4927444} {\bibfield
  {journal} {\bibinfo  {journal} {Applied Physics Letters}\ }\textbf {\bibinfo
  {volume} {107}},\ \bibinfo {pages} {062601} (\bibinfo {year}
  {2015})}\BibitemShut {NoStop}%
\bibitem [{\citenamefont {Gao}\ \emph {et~al.}(2006)\citenamefont {Gao},
  \citenamefont {Zmuidzinas}, \citenamefont {Mazin}, \citenamefont {Day},\ and\
  \citenamefont {Leduc}}]{gaoExperimentalStudyKinetic2006}%
  \BibitemOpen
  \bibfield  {author} {\bibinfo {author} {\bibfnamefont {J.}~\bibnamefont
  {Gao}}, \bibinfo {author} {\bibfnamefont {J.}~\bibnamefont {Zmuidzinas}},
  \bibinfo {author} {\bibfnamefont {B.~A.}\ \bibnamefont {Mazin}}, \bibinfo
  {author} {\bibfnamefont {P.~K.}\ \bibnamefont {Day}}, \ and\ \bibinfo
  {author} {\bibfnamefont {H.~G.}\ \bibnamefont {Leduc}},\ }\bibfield  {title}
  {\enquote {\bibinfo {title} {Experimental study of the kinetic inductance
  fraction of superconducting coplanar waveguide},}\ }\href {\doibase
  10.1016/j.nima.2005.12.075} {\bibfield  {journal} {\bibinfo  {journal}
  {Nuclear Instruments and Methods in Physics Research Section A: Accelerators,
  Spectrometers, Detectors and Associated Equipment}\ }\bibinfo {series}
  {Proceedings of the 11th {{International Workshop}} on {{Low Temperature
  Detectors}}},\ \textbf {\bibinfo {volume} {559}},\ \bibinfo {pages}
  {585--587} (\bibinfo {year} {2006})}\BibitemShut {NoStop}%
\bibitem [{\citenamefont
  {{Castellanos-Beltran}}(2010)}]{castellanos-beltranDevelopmentJosephsonParametric2010}%
  \BibitemOpen
  \bibfield  {author} {\bibinfo {author} {\bibfnamefont {M.~A.}\ \bibnamefont
  {{Castellanos-Beltran}}},\ }\emph {\bibinfo {title} {Development of a
  {{Josephson}} Parametric Amplifier for the Preparation and Detection of
  Nonclassical States of Microwave Fields}},\ \href@noop {} {Ph.D. thesis},\
  \bibinfo  {school} {University of Colorado}, \bibinfo {address} {{Boulder}}
  (\bibinfo {year} {2010})\BibitemShut {NoStop}%
\bibitem [{\citenamefont {Rauscher}\ \emph {et~al.}(2016)\citenamefont
  {Rauscher}, \citenamefont {Janssen},\ and\ \citenamefont
  {Minihold}}]{rauscherFundamentalsSpectrumAnalysis2016a}%
  \BibitemOpen
  \bibfield  {author} {\bibinfo {author} {\bibfnamefont {C.}~\bibnamefont
  {Rauscher}}, \bibinfo {author} {\bibfnamefont {V.}~\bibnamefont {Janssen}}, \
  and\ \bibinfo {author} {\bibfnamefont {R.}~\bibnamefont {Minihold}},\
  }\href@noop {} {\emph {\bibinfo {title} {Fundamentals of Spectrum
  Analysis}}},\ \bibinfo {edition} {9th}\ ed.\ (\bibinfo  {publisher} {{Rohde
  \& Schwarz}},\ \bibinfo {address} {{M{\"u}nchen}},\ \bibinfo {year} {2016})\
  \bibinfo {note} {oCLC: 964521293}\BibitemShut {NoStop}%
\bibitem [{\citenamefont {Planat}\ \emph {et~al.}(2019)\citenamefont {Planat},
  \citenamefont {Dassonneville}, \citenamefont {Mart{\'i}nez}, \citenamefont
  {Foroughi}, \citenamefont {Buisson}, \citenamefont {{Hasch-Guichard}},
  \citenamefont {Naud}, \citenamefont {Vijay}, \citenamefont {Murch},\ and\
  \citenamefont {Roch}}]{planatUnderstandingSaturationPower2019}%
  \BibitemOpen
  \bibfield  {author} {\bibinfo {author} {\bibfnamefont {L.}~\bibnamefont
  {Planat}}, \bibinfo {author} {\bibfnamefont {R.}~\bibnamefont
  {Dassonneville}}, \bibinfo {author} {\bibfnamefont {J.~P.}\ \bibnamefont
  {Mart{\'i}nez}}, \bibinfo {author} {\bibfnamefont {F.}~\bibnamefont
  {Foroughi}}, \bibinfo {author} {\bibfnamefont {O.}~\bibnamefont {Buisson}},
  \bibinfo {author} {\bibfnamefont {W.}~\bibnamefont {{Hasch-Guichard}}},
  \bibinfo {author} {\bibfnamefont {C.}~\bibnamefont {Naud}}, \bibinfo {author}
  {\bibfnamefont {R.}~\bibnamefont {Vijay}}, \bibinfo {author} {\bibfnamefont
  {K.}~\bibnamefont {Murch}}, \ and\ \bibinfo {author} {\bibfnamefont
  {N.}~\bibnamefont {Roch}},\ }\bibfield  {title} {\enquote {\bibinfo {title}
  {Understanding the {{Saturation Power}} of {{Josephson Parametric Amplifiers
  Made}} from {{SQUID Arrays}}},}\ }\href {\doibase
  10.1103/PhysRevApplied.11.034014} {\bibfield  {journal} {\bibinfo  {journal}
  {Physical Review Applied}\ }\textbf {\bibinfo {volume} {11}},\ \bibinfo
  {pages} {034014} (\bibinfo {year} {2019})}\BibitemShut {NoStop}%
\bibitem [{\citenamefont {Gely}\ and\ \citenamefont
  {Steele}(2019)}]{gelyQuCATQuantumCircuit2019}%
  \BibitemOpen
  \bibfield  {author} {\bibinfo {author} {\bibfnamefont {M.~F.}\ \bibnamefont
  {Gely}}\ and\ \bibinfo {author} {\bibfnamefont {G.}~\bibnamefont {Steele}},\
  }\bibfield  {title} {\enquote {\bibinfo {title} {{{QuCAT}}: {{Quantum Circuit
  Analyzer Tool}} in {{Python}}},}\ }\href {\doibase 10.1088/1367-2630/ab60f6}
  {\bibfield  {journal} {\bibinfo  {journal} {New Journal of Physics}\ }
  (\bibinfo {year} {2019}),\ 10.1088/1367-2630/ab60f6}\BibitemShut {NoStop}%
\bibitem [{\citenamefont {Noschese}\ \emph {et~al.}(2013)\citenamefont
  {Noschese}, \citenamefont {Pasquini},\ and\ \citenamefont
  {Reichel}}]{noscheseTridiagonalToeplitzMatrices2013}%
  \BibitemOpen
  \bibfield  {author} {\bibinfo {author} {\bibfnamefont {S.}~\bibnamefont
  {Noschese}}, \bibinfo {author} {\bibfnamefont {L.}~\bibnamefont {Pasquini}},
  \ and\ \bibinfo {author} {\bibfnamefont {L.}~\bibnamefont {Reichel}},\
  }\bibfield  {title} {\enquote {\bibinfo {title} {Tridiagonal {{Toeplitz}}
  matrices: Properties and novel applications},}\ }\href {\doibase
  10.1002/nla.1811} {\bibfield  {journal} {\bibinfo  {journal} {Numerical
  Linear Algebra with Applications}\ }\textbf {\bibinfo {volume} {20}},\
  \bibinfo {pages} {302--326} (\bibinfo {year} {2013})}\BibitemShut {NoStop}%
\bibitem [{\citenamefont {Nigg}\ \emph {et~al.}(2012)\citenamefont {Nigg},
  \citenamefont {Paik}, \citenamefont {Vlastakis}, \citenamefont {Kirchmair},
  \citenamefont {Shankar}, \citenamefont {Frunzio}, \citenamefont {Devoret},
  \citenamefont {Schoelkopf},\ and\ \citenamefont
  {Girvin}}]{niggBlackBoxSuperconductingCircuit2012}%
  \BibitemOpen
  \bibfield  {author} {\bibinfo {author} {\bibfnamefont {S.~E.}\ \bibnamefont
  {Nigg}}, \bibinfo {author} {\bibfnamefont {H.}~\bibnamefont {Paik}}, \bibinfo
  {author} {\bibfnamefont {B.}~\bibnamefont {Vlastakis}}, \bibinfo {author}
  {\bibfnamefont {G.}~\bibnamefont {Kirchmair}}, \bibinfo {author}
  {\bibfnamefont {S.}~\bibnamefont {Shankar}}, \bibinfo {author} {\bibfnamefont
  {L.}~\bibnamefont {Frunzio}}, \bibinfo {author} {\bibfnamefont {M.~H.}\
  \bibnamefont {Devoret}}, \bibinfo {author} {\bibfnamefont {R.~J.}\
  \bibnamefont {Schoelkopf}}, \ and\ \bibinfo {author} {\bibfnamefont {S.~M.}\
  \bibnamefont {Girvin}},\ }\bibfield  {title} {\enquote {\bibinfo {title}
  {Black-{{Box Superconducting Circuit Quantization}}},}\ }\href {\doibase
  10.1103/PhysRevLett.108.240502} {\bibfield  {journal} {\bibinfo  {journal}
  {Physical Review Letters}\ }\textbf {\bibinfo {volume} {108}},\ \bibinfo
  {pages} {240502} (\bibinfo {year} {2012})}\BibitemShut {NoStop}%
\bibitem [{\citenamefont {Vool}\ and\ \citenamefont
  {Devoret}(2017)}]{vool_introductionquantum_2017}%
  \BibitemOpen
  \bibfield  {author} {\bibinfo {author} {\bibfnamefont {U.}~\bibnamefont
  {Vool}}\ and\ \bibinfo {author} {\bibfnamefont {M.}~\bibnamefont {Devoret}},\
  }\bibfield  {title} {\enquote {\bibinfo {title} {Introduction to quantum
  electromagnetic circuits: {{Introduction}} to quantum electromagnetic
  circuits},}\ }\href {\doibase 10.1002/cta.2359} {\bibfield  {journal}
  {\bibinfo  {journal} {International Journal of Circuit Theory and
  Applications}\ } (\bibinfo {year} {2017}),\ 10.1002/cta.2359}\BibitemShut
  {NoStop}%
\bibitem [{\citenamefont {Simons}(2001)}]{simonsCoplanarWaveguideCircuits2001}%
  \BibitemOpen
  \bibfield  {author} {\bibinfo {author} {\bibfnamefont {R.~N.}\ \bibnamefont
  {Simons}},\ }\href {\doibase 10.1002/0471224758} {\emph {\bibinfo {title}
  {Coplanar {{Waveguide Circuits}}, {{Components}}, and {{Systems}}}}},\ edited
  by\ \bibinfo {editor} {\bibfnamefont {K.}~\bibnamefont {Chang}},\ Wiley
  {{Series}} in {{Microwave}} and {{Optical Engineering}}\ (\bibinfo
  {publisher} {{John Wiley \& Sons, Inc.}},\ \bibinfo {address} {{New York,
  USA}},\ \bibinfo {year} {2001})\BibitemShut {NoStop}%
\bibitem [{\citenamefont {Schmidt}\ \emph {et~al.}()\citenamefont {Schmidt},
  \citenamefont {Bothner}, \citenamefont {Rodrigues}, \citenamefont {Gely},
  \citenamefont {Jenkins},\ and\ \citenamefont {Steele}}]{zenodo1}%
  \BibitemOpen
  \bibfield  {author} {\bibinfo {author} {\bibfnamefont {F.~E.}\ \bibnamefont
  {Schmidt}}, \bibinfo {author} {\bibfnamefont {D.}~\bibnamefont {Bothner}},
  \bibinfo {author} {\bibfnamefont {I.~C.}\ \bibnamefont {Rodrigues}}, \bibinfo
  {author} {\bibfnamefont {M.~F.}\ \bibnamefont {Gely}}, \bibinfo {author}
  {\bibfnamefont {M.~D.}\ \bibnamefont {Jenkins}}, \ and\ \bibinfo {author}
  {\bibfnamefont {G.~A.}\ \bibnamefont {Steele}},\ }\bibfield  {title}
  {\enquote {\bibinfo {title} {Data and processing for ``{{Current}} detection
  using a {{Josephson}} parametric upconverter''},}\ }\href {\doibase DOI} {\
  DOI}\BibitemShut {NoStop}%
\end{thebibliography}
\end{document}